%% file: main.tex
\documentclass[preprint,3p,12pt]{elsarticle}
\journal{arXiv}

\usepackage{psfrag}
\usepackage{graphicx}
\usepackage{paralist}
\usepackage{color}
\usepackage{amsmath}
\usepackage{amssymb}
\usepackage{varioref}
\usepackage{fancyhdr}
\usepackage{amsfonts}
\usepackage{bm}
\usepackage{float}
\usepackage{booktabs}
\usepackage{multirow}
\usepackage{array}
\usepackage{mathptmx}
\usepackage{natbib}
\usepackage{epstopdf}
\usepackage{threeparttable}
\usepackage[numbers]{natbib}
\usepackage{enumitem}
\usepackage{tabularx}
\usepackage{caption,subcaption}

\input{macros}

\begin{document}

\title{An Artificial-intelligence/Statistics Solution to Quantify Material Distortion for Thermal Compensation in Additive Manufacturing}



\author[1]{Chao Wang}

\author[1]{Shaofan Li\corref{cor1}}

\author[2]{Danielle Zeng}

\author[3]{Xinhai Zhu}

\cortext[cor1]{Corresponding Author, Email:~shaofan@berkeley.edu}
\address[1]{Department of Civil and Environmental Engineering, University of California, Berkeley,
CA 94720, USA}
\address[2]{Ford Motor Co., Detroit, MI 48226, USA}
\address[3]{ANSYS Livermore Software Technology (ANSYS-LST), Livermore, CA94551, USA }


\begin{abstract}
In this paper, we introduce a probabilistic statistics solution
or artificial intelligence (AI) approach to identify and quantify
permanent (non-zero strain) continuum/material deformation
only based on the scanned material data in the spatial configuration
and the shape of the initial design configuration or
the material configuration.
The challenge of this problem is that
we only know the scanned material data in the spatial configuration
and the shape of the design configuration of three-dimensional (3D)
printed products, whereas for a specific scanned material point
we do not know its corresponding material coordinates in the initial or designed
referential configuration, provided that we do not know the detailed information
on actual physical deformation process.
Different from physics-based modeling, the method
developed here is a data-driven artificial intelligence method, which
solves the problem with incomplete deformation data or with
missing information of actual physical deformation process.
We coined the method is an
AI-based {\it material deformation finding} algorithm.

This method has practical significance and important applications
in finding and designing thermal compensation configuration of
a 3D printed product in additive manufacturing, which is at the heart of
the cutting edge 3D printing technology.
In this paper,
we demonstrate that the proposed AI continuum/material deformation finding approach
can accurately find permanent thermal deformation configuration
for a complex 3D printed structure component,
and hence to identify the thermal compensation design configuration
in order to minimizing the impact of temperature fluctuations on 3D printed
structure components that are sensitive to changes of temperature.
\end{abstract}




\maketitle

\section{Introduction}
Due to some of its great technological advantages,
Additive manufacturing (AM) has been an international hot-spot in the past several years,
such as unparalleled design freedom and short lead time \cite{herzog2016additive}.
The unique edge of three-dimensional (3D) printing technology
lies in its ability to directly fabricate complex shapes
and eliminate the waste that are associated with the conventional manufacturing \cite{campbell2011could}.
In 3D printing manufacture process, geometric complexity has no influence on building efficiency,
and therefore, no additional effort is required to design the molding structure,
which makes 3D printing a very promising manufacturing technique \cite{mueller2012additive,wong2012review}.

Despite these promising features,
one of the main challenges for large-scale industrial use of 3D printing technology
is how to eliminate or mitigate the structural distortion due to thermal effect during
3D printing processes,
because it is a direct concern going against the main advantages of 3D printing: precision and as a single
step manufacture.
A typical 3D printing process is usually involved with melting materials into liquid or mushy-state
and then rapidly prototyping it into various designed shapes. This process is a
temperature drastic change process, and it causes material shrinkage,
and various thermal-induced permanent deformations
as well as residual stresses, which are induced
when temperature
changes during the printing process.
This is not a simple thermal-mechanical process, and it is
a complex interaction in and between the material layers in
different material phases,
and it governed by underlying process parameter variations and
fluctuations \cite{wang1996influence}.

\begin{figure}[t]
\begin{center}
\includegraphics[width=4.5in]{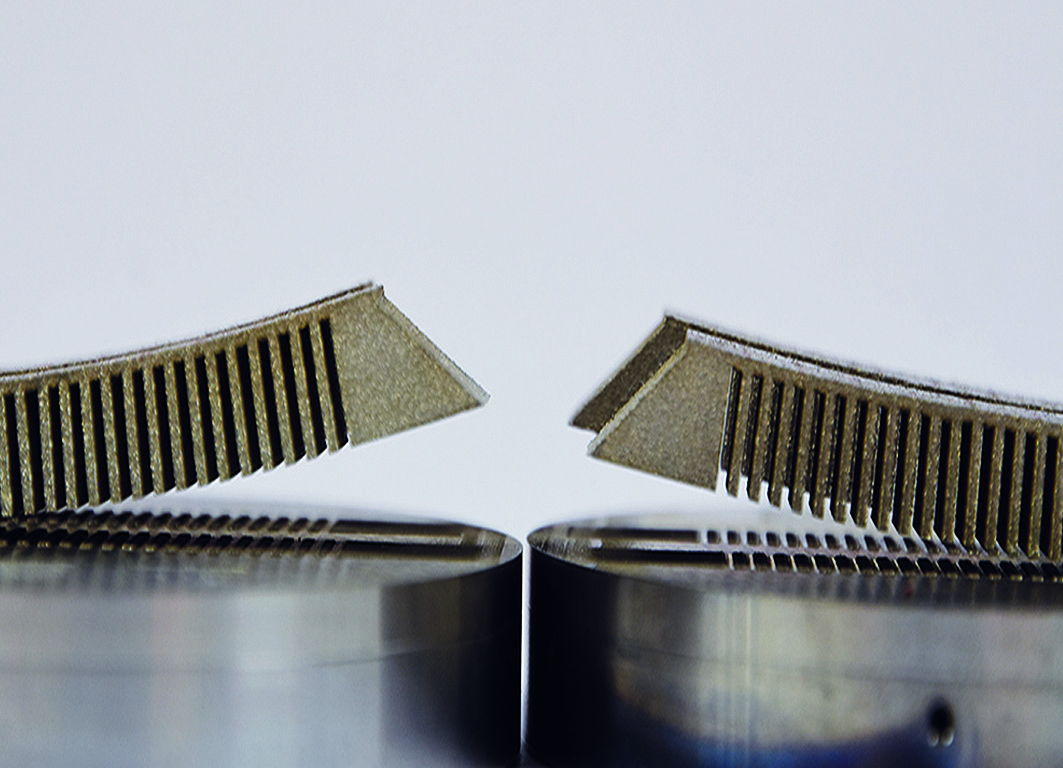}
\end{center}
\caption{An example of thermal distorted 3D printed product:
Distortion on a 3D printed Inconel competent heated only
from below (left) versus the same Inconel component that was
made when heated from above and below (right).
(From \underline{https://www.pinterest.com/pin/411305378465700906/
})
}
\label{Example of Distortion}
\end{figure}

In order to address this issue, various approaches
or technologies have been used or proposed for distortion mitigation
or compensation. The key for deformation compensation is to
accurately predict deformation compensation allowance
under a given set of 3D printing process parameters.
It turns out that this is a very challenging problem that is largely open up to date.
The most approaches to this problem are employing
physics-based modeling to estimate the distortion allowance based
on simulations of the actual physical process of 3D printing.
For example, the Finite Element Analysis (FEA) has been implemented to computer
residual stresses and thermal distortions in various additive manufacturing settings,
e.g. \cite{mukherjee2017improved,arrazola2013recent,schoinochoritis2017simulation,yin2012simulation}.
By using coupled thermal, fluid flow and structure finite element models,
researchers have studied
the evolution of stresses and strains in various 3D printed
products
and their dependence on major process parameters including heat input and layer thickness,
etc. \cite{mukherjee2017improved}.
However,
these FEA modelings heavily depend on the accuracy and fidelity of the physical model adopted,
and they are often limited by our lack of understanding of the actual physical process
of a complex 3D printing process under investigation.
Moreover, it is also very
challenging to acquire the vast number of model parameters accurately in a actual manufacture process,
and, on the other hand, the sophistication of the physical model also decreases its convenience
to simulate the additive manufacturing process in order to compensate shape precisely.

In recent years,
a number of data-driven methods in additive manufacturing have been developed
based on existing machine learning strategies.
For example,
a novel data-driven approach that uses multi-scale multi-physics material models
for additive manufacturing by self-consistent clustering analysis (SCA)
has been developed in \cite{yan2018data}.
Li et al. \cite{li2018integration} proposed
an integrated framework that includes both classical numerical simulation and statistical method, such as,
Gaussian process, and Bayesian calibration to predict thermal-mechanical field in additive manufacturing.
These data-driven approaches
use machine learning techniques to supplement the physics-based numerical modeling and simulation solutions.
So far, to the best of authors' knowledge, there has been little discussion, if there is any,
on a fully data-driven approach in prediction of deformation or thermal
compensation for additive manufacturing.

In this work, we have developed an entirely new data-driven approach
to identify and quantify continuum/material distortion field in a 3D printed product
in additive manufacturing. Our solution is solely based on
the scan data on the spatial position of material points of the printed product
and the designed configuration or designed data of the desired product, i.e. the initial or material configuration.
The proposed continuum/material deformation finding algorithm or approach is closely interlinked to the coherent point drift (CPD) algorithm in artificial intelligence and statistics
\cite{myronenko2010point}, and it may be viewed as
a refined CPD algorithm for continuum deformation identification.
The original CPD algorithm has demonstrated its dominance over most leading edge discrete point set
registration methods. Nevertheless, to the best of authors' knowledge,
how to apply it to 3D printing technology has not been studied.
On the other hand,
the proposed material finding algorithm is a further development and an extension of original CPD algorithm
to find continuum deformation or material registration in continuum media by
aligning two stereolithography (STL) files.

There are five sections in this paper as follows:
We start by recalling some key elements of original CPD algorithm in Section 2.
Section 3 describes the proposed continuum/material deformation finding algorithm featuring the technique
of aligning STL files.
Numerical examples as well as the comparison of the numerical results
with that of the CPD algorithm are provided in section 4.
Finally, Section 5 provides the  conclusion of this paper with summary.

\section{Review of the Coherent Point Drift (CPD) algorithm}
Since the resemblance between the {\it continuum/material deformation finding algorithm}
proposed in this work and the original Coherent Point Drift (CPD) algorithm
\cite{Myronenko2007,myronenko2010point,Wang2011},
for sake of self-containedness, we first briefly review the
CPD method for non-rigid point registration such that the readers can become
familiar with the idea, the principle, and the main techniques of the method
and relevant literatures.

\begin{figure}[ht]
\begin{subfigure}{.5\textwidth}
  \centering
  \includegraphics[width=.9\linewidth]{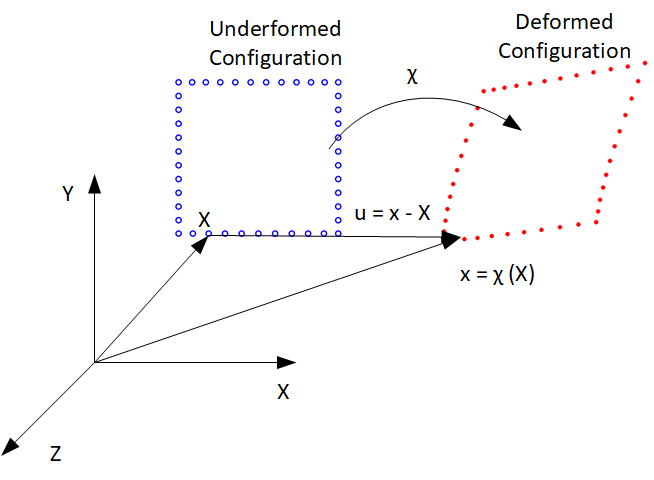}
  \caption{}
  \label{fig:sub-first}
\end{subfigure}
\begin{subfigure}{.5\textwidth}
  \centering
  \includegraphics[width=.9\linewidth]{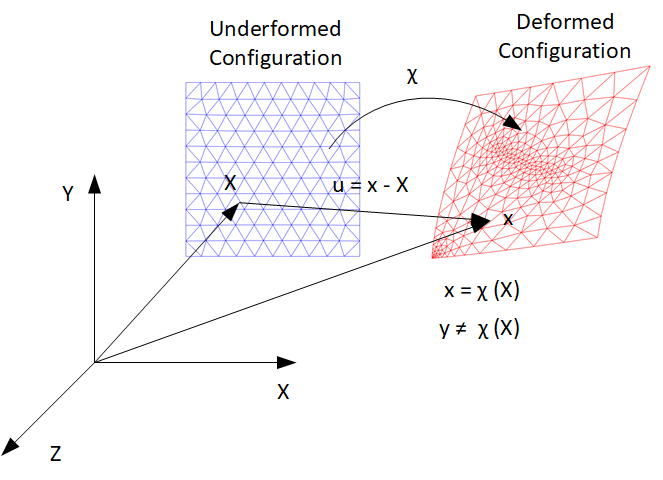}
  \caption{}
\end{subfigure}
\caption{ The comparison between non-rigid registration method and material deformation finding method:
(a) Non-rigid registration problem in computer vision vs.
(b) Material deformation finding problem in 3D printing and continuum mechanics .}
\label{fig:boudnary vs dense}
\end{figure}

Originally, the Coherent Point Drift (CPD) method
was first proposed and developed by Myronenko and Song in \cite{myronenko2010point}.
An example of registration problem which could be solved
by CPD algorithm is illustrated in Fig. \ref{fig:boudnary vs dense}(a).
The CPD algorithm first
assumes that two point sets are given:
(1) $\mathbf{x}_{m \times D}=\left(\mathbf{x}_{1}, \ldots, \mathbf{x}_{m}\right)^{T}$
is the first set of data for spatial point positions in deformation configuration, and
(2)
$\mathbf{X}_{N \times D}=\left(\mathbf{X}_{1}, \ldots, \mathbf{X}_{N}\right)^{T}$
is the second set of point positions in un-deformed configuration.
After having those two point sets,
the goal of this registration problem is twofold:
(1) Align two point sets and find the correspondence pairs of points
$(\mathbf{X}_I,\mathbf{x}_j)$, and
(2) Find the non-uniform deformation map $\chi$.

CPD algorithm cast the point set registration into a Maximum Likelihood Estimation (MLE) problem.
For example, one can choose $\mathbf{X}_{N \times D}$  as the data points
and $\mathbf{x}_{m \times D}$ as the GMM centroid.
By considering a statistical MLE or optimization,
in which one maximizes the  probability for each data point by moving the GMM centroid,
the spatial points may move to align (overlap) with the second set of points, i.e
data points, in the undeformed configuration.
That is to say, at the optimum, we match the points
in two sets and obtain the correspondence pairs of points $(\mathbf{X}_I,\mathbf{x}_i)$.
Finally, as in continuum mechanics, the non-uniform deformation map $\chi: \mathbb{R}^D \rightarrow \mathbb{R}^D$ could be defined by the corresponding pairs  $(\mathbf{X}_I,\mathbf{x}_i)$ with parameter $\theta$,
\begin{equation}
    \mathbf{x}_j = \chi(\mathbf{X}_I,\theta)
\end{equation}

The marginal  probability density function for ${\bf X}_I$
used in the CPD algorithm is defined as follows,
\begin{equation}
p({\bf X}_I )=w \frac{1}{N}+(1-w) \sum_{j=1}^{m} p({\bf x}_j) p(\mathbf{X}_I | {\bf x}_j),
\label{eq:GMM}
\end{equation}
where $j=1, 2, \cdots, m$
($j=m+1$ is an added degree of freedom)
is the index for the GMM centroid, and the index $I=1,2, \cdots , N$ is
for the data points. $p({\bf X}_I |{\bf x}_j)$ are conditional probability of data points with
respect to each Gaussian distribution, while
$P({\bf x}_j)$ are  membership probability for ${\bf x}_j$.
Here we assume that each component probability shares the same value,
\begin{equation}
  P({\bf x}_j) = \frac{1}{m}.
\end{equation}
In order to taking into account for noises and outliers, a uniform probability mass
function $P({\bf x}_{m+1})= \displaystyle \frac{1}{N}$
is adopted for the mixture model.
Then the final density function is a mixture model of
original GMM and uniform distribution with parameter $w$
which shows our prior belief of the level of noise in the data.

In the original CPD algorithm, the covariances matrix $\Sigma_j$ for each Gaussian distribution
are assume to be the same,
\begin{equation}
    \Sigma_1 = \Sigma_2= ... \Sigma_j = \sigma^2 \cdot \mathbb{I}
\end{equation}
where $\mathbb{I}$ is the identical matrix and all covariances matrix $\Sigma_j$
are controlled by the variance $\sigma^2$.
Then the conditional probability density function of $I$-th data point ${\bf X}_I$
on $j$-th component in GMM is,
\begin{equation}
 p(\mathbf{X}_I | {\bf x}_j)=\frac{1}{\left(2 \pi \sigma^{2}\right)^{D / 2}}
 \exp \Bigl( {-\frac{\left\| {\bf X}_I- {\bf x}_j \right\| 2}{2 \sigma^{2}}}
 \Bigr)~,
\end{equation}
where the space is $D$-dimensional, in the registration cases, usually $D$ equals 2 or 3.

In non-uniform deformation, the new position ${\bf x}_j'$ of a GMM centroid ${\bf x}_j$ is defined by the inverse of  mapping function $\chi^{-1}$ with parameter $\theta$,
\begin{equation}
{\bf x}_j'=\chi^{-1}({\bf x}_j,\theta)
\end{equation}
Thus, in essence, the CPD method is to use
the maximum likelihood estimate to update the spatial position of GMM centroid ${\bf x}=\{ {\bf x}_j \}_{j=1}^m$ and covariance
$\sigma^2$
such that it can converge to the data point set $\mathbf{X}$.
For instance, we may define the likelihood function as,
\begin{equation}
\mathcal{L} ({\bf X};{\bf x},\theta, \sigma^{2}) = \Pi_{I=1}^N   p ({\bf X}_I;{\bf x},\theta, \sigma^{2})
\label{eq:GMM2}
\end{equation}
that is
\begin{equation}
( \hat{\theta}, \hat{\sigma}^2)
= \underset{{\bf \theta ,\sigma^2}}{\operatorname{argmax}} ~\log \Bigl( \mathcal{L}({\bf X};{\bf x},\theta, \sigma^{2}) \Bigr)
\end{equation}
where $\hat{\theta}, \hat{\sigma}^2$ are the optimal values which
determine both the GMM distribution and deformation mapping $\chi$.

In actual computations, in order to find the point sets correspondence,
CPD formulations use a so-called expectation-maximization (EM) algorithm,
\cite{dempster1977maximum,bishop1995neural},
which is an iterative or alternating method
to find the maximum likelihood estimators
on various incomplete data in statistics.
More precisely speaking,
the EM algorithm can use an iterative solution
to find the maximum likelihood estimate or maximum a posteriori (MAP)
estimate of interested parameters,
where the statistics model heavily depends on missing data
or the so-called latent variables.
In EM algorithm, the first iterative step
is a soft imputation on the missing data with
a given probability distribution
for all its possible values (E-step),
and then the second step is to
update parameters with the given imputed data (M-step)
by maximizing or minimizing objective function $Q$ that is
found in the E-step.
If we focus on the system parameters  $\sigma^{2}$ and $\theta$, the objective function $Q$ is defined as,
\begin{equation}
\mathcal{L}_Q \left(\theta, \sigma^{2}\right)=\frac{1}{2 \sigma^{2}} \sum_{j=1}^{m} \sum_{I=1}^{N}
p^{\mathrm{old}}\left({\bf X}_I | {\bf x}_{j}\right)\left\| {\bf X}_{I}-
\chi^{-1}\left( {\bf x}_{j}, \theta\right)\right\|^{2}+\frac{N_{p} D}{2}
\log \sigma^{2},
\label{eq:Q function}
\end{equation}
where ${\bf x}_{j}' = \chi^{-1}\left( {\bf x}_{j}, \theta\right)$ is the transformed new GMM centroid locations;
\[
N_{p}=\sum_{j=1}^{m} \sum_{I=1}^{N} p^{\text {old }}\left({\bf X}_I | {\bf x}_{j}\right)
\]
and
\begin{equation}
p^{\text {old }}\left({\bf X}_I | {\bf x}_{j}\right)=
\frac{\exp \left(-\frac{1}{2}\left\|\frac{{\bf X}_{I}-\chi^{-1}\left( {\bf x}_{j},
\theta^{\text {old }}\right)}{\sigma^{\text {old }}}\right\|^{2}\right)}
{\sum_{j=1}^{M} \exp \left(-\frac{1}{2}\left\|\frac{ {\bf X}_{I}-\chi^{-1}\left({\bf x}_{j},
 \theta^{\text {old }}\right)}{\sigma^{\text {old }}}
\right\|^{2}\right)+\left(2 \pi \sigma^{2}\right)^{D / 2} \frac{w}{1-w} \frac{m}{N}}~.
\end{equation}

Another way to think about EM algorithm is, after obtaining the system parameters
$\hat{\theta}, \hat{\sigma}^2$ in E-step, we may either move ${\bf X}$ to
the deformed configuration, or move ${\bf x}$ to the undeformed configuration.
By assuming that  ${\bf X}^{n+1} = {\bf X}^n + {\bf V}^n \Delta t$ or
 ${\bf x}^{n+1} = {\bf x}^n + {\bf v}^n \Delta t$, where $\Delta t$ is just a scalar parameter.
 The above objective function is also a function of the velocity field, i.e.
 $\mathcal{L} ({\bf X};{\bf x},\theta, \sigma^{2}, {\bf V}) $.
 Thus, we may perform a maximization (M-step) of the objective function, i.e.
 \begin{equation}
\hat{\bf V}
= \underset{{\bf \theta ,\sigma^2}}{\operatorname{argmax}} ~\log \Bigl( \mathcal{L}({\bf X};{\bf x},\theta, \sigma^{2}, {\bf V}) \Bigr)
\end{equation}
Once we obtain the velocity field ${\bf V}$, we find the new position of GMM centroid, i.e. ${\bf X}^{n+1} = {\bf X}^n + {\bf V}^n \Delta t$, where ${\bf V}^n = \hat{\bf V}$.
Once this is accomplished, we come back to estimate $\theta$ and $\sigma^2$ again, and subsequently ${\bf V}^{n+1}$ and
${\bf X}^{n+1}$, until the convergence of the two point sets, i.e. ${\bf X} \to {\bf x}$.

The essence of CPD algorithm is to ensure that the topological structure of data sets are preserved during registration.
In order to address some computational issues of coherent moving,
we may need add one prior term $\mathcal{R}$ to the MLE formulation
and replace it into a Maximum a posteriori (MAP) estimation.
The prior term encoded our prior belief of moving coherent and smoothness of displacement field.
To do so, one can modify Eq. (\ref{eq:Q function})
by adding a regularization term $\mathcal{R}$ to obtain a constrained optimization problem,
\begin{equation}
\mathcal{L}_{Q\mathcal{R}}\left(\theta, \sigma^{2}\right)=\mathcal{L}_Q\left(\theta, \sigma^{2}\right)+\mathcal{R}
\left(\mathcal{T}\left(\mathbf{x}, \theta\right)\right),
\end{equation}
Specifically, the regularization term used in nonrigid CPD algorithms is
\begin{equation}
\mathcal{R}\left(\mathcal{T}\left(\mathbf{x},
\theta\right)\right)
=\int_{\mathbb{R}^{D}} \frac{|\tilde{v}(\mathbf{s})|^{2}}{\tilde{G}(\mathbf{s})} d \mathbf{s}
\label{eq:regularization}
\end{equation}
where ${G}_{m \times m}$ is the Gaussian kernel matrix with element $g_{i j}=\exp \left(-\frac{1}{2}\left\|\frac{\mathbf{x}_{i}-\mathbf{x}_{j}}{\beta}\right\|^{2}\right)$, $\tilde{{G}}$ is its Fourier transformation. $\tilde{v}$ is the Fourier transformation of the velocity field.
The integration is taken in the frequency domain with variable $\mathbf{s}$.
This regularization term can first filter the high frequency part of the displacement field and then add penalty on it. By penalty the high frequency energy, this regularization term force the displacement field to put more energy on the low frequency domain and finally smooth the displacement field.

\begin{figure}[h]
     \centering
     \begin{minipage}{0.49\linewidth}
         \centering
         \includegraphics[width=2.7in]{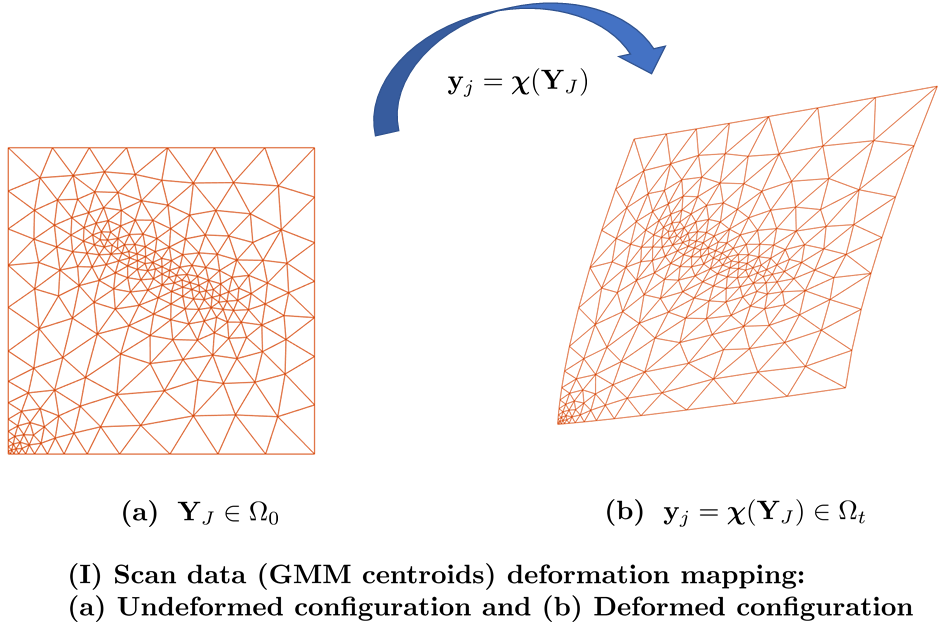}
     \end{minipage}
     \begin{minipage}{0.49\linewidth}
         \centering
         \includegraphics[width=2.7in]{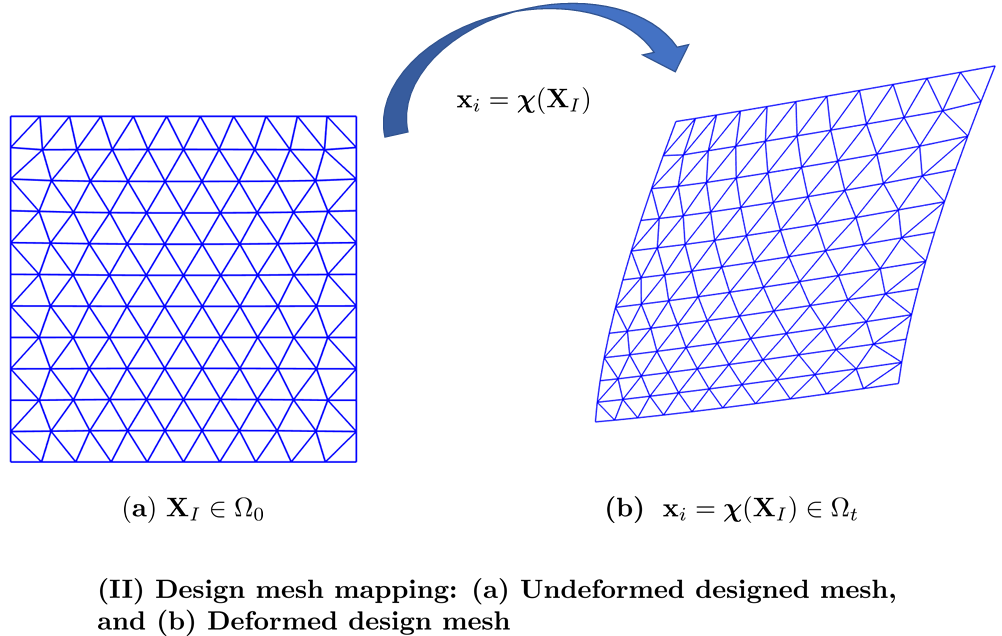}
     \end{minipage}\\
     \bigskip
     \begin{minipage}{0.49\linewidth}
         \centering
         \includegraphics[width=2.9in]{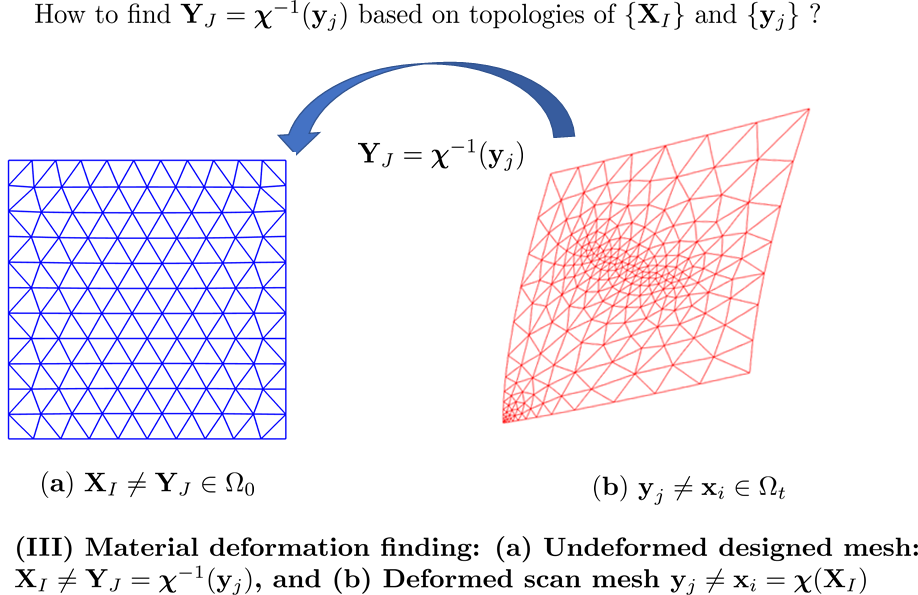}
     \end{minipage}
      \begin{minipage}{0.49\linewidth}
         \centering
         \includegraphics[width=2.7in]{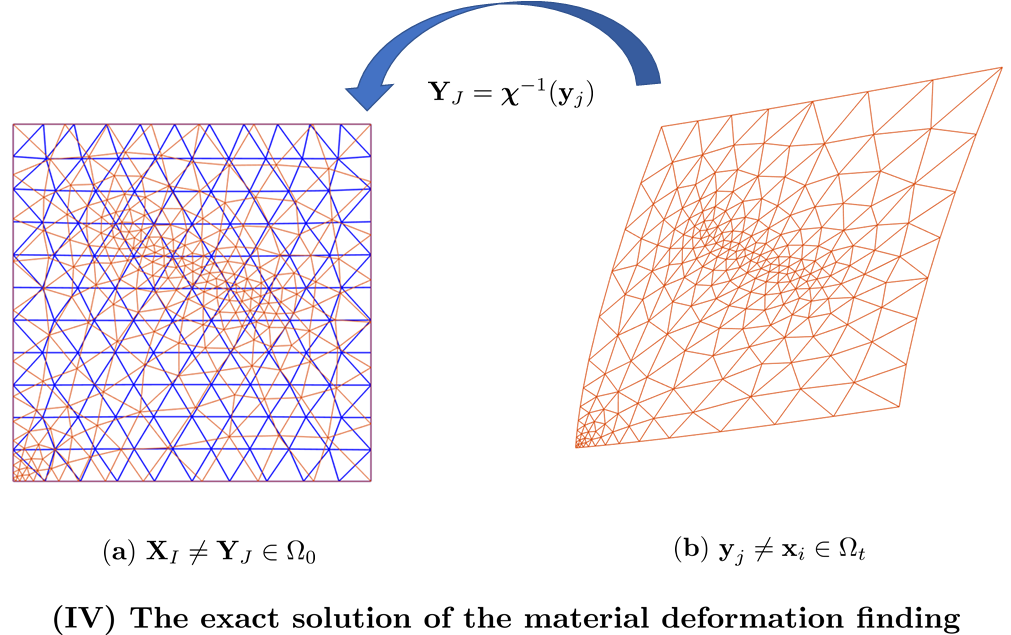}
     \end{minipage}
        \caption{Material deformation finding for a deformed two-dimensional unit square:
        (I) Deformation of scanned data from the initial configuration to deformed configuration;
        (II) Deformation of designed mesh; (III) Topology connection between undeformed design mesh
        and scanned data in deformed configuration, and
        (IV) The exact solution for the material deformation finding problem.}
        \label{fig:mesh}
\end{figure}

\section{The material registration in continuum media}

\subsection{Comparison between CPD and material deformation finding algorithm}
In this work, a material registration in continuum media or a
material deformation finding algorithm is developed to match the designed mesh of
a 3D printed model and scanned material data or mesh of the final printed product.
We also
assume that two point sets are given, as shown in Fig. \ref{fig:boudnary vs dense}(b):
(1) $\mathbf{y}_{m \times D}=\left(\mathbf{y}_{1}, \ldots, \mathbf{y}_{m}\right)^{T}$
is the first set of data for spatial point positions in deformation configuration, and
(2)
$\mathbf{X}_{N \times D}=\left(\mathbf{X}_{1}, \ldots, \mathbf{X}_{N}\right)^{T}$
is the second set of point positions in un-deformed configuration.
It should be noted that
the spatial point set is denoted by $\mathbf{y}$ instead of $\mathbf{x}$ in material deformation finding algorithm. The reason for this modification will be demonstrated in the following paragraph.

The main challenge in this method is that there is no one to one correspondence
between the nodes in two sets of meshes. A partial correspondence on the surface of the product
may be established or identified by some geometric object identification methods, however,
this can only be done for a group of nodes.
In other words, even though we know the true deformation mapping function $\chi$, we still could not find the image of a material point $\mathbf{X}_I$ in the spatial point set $\mathbf{y}$,
\begin{equation}
    \mathbf{y}_j \neq \chi(\mathbf{X}_I), ~~
    \forall j = 1,2,\cdots,m
\end{equation}
Therefore, the material deformation finding algorithm is to figure out the deformation mapping function $\chi$ between two point sets $\mathbf{X}$ and $\mathbf{y}$ without the help of correspondence.

Besides the main difference above, there are also another three important differences between point set registration and material deformation finding problem.
Firstly, in computer vision problem, both feature points in two point sets are obtained by feature extraction form image or triangulation data which means that each feature sets may contain noise. Our material deformation finding problem, however, use the exact design mesh as one feature set. Therefore, only the scan data set may contain noise data and the design data set is noise-free.

Second, in most cases, the feature points in computer vision application are only located on the boundary of the given model. On the contrary, the material deformation finding problem, as shown in Fig. \ref{fig:boudnary vs dense}, has feature points not only on the boundary but also in the interior domain.

\begin{figure}[ht]
\begin{subfigure}{.5\textwidth}
  \centering
  \includegraphics[width=.9\linewidth]{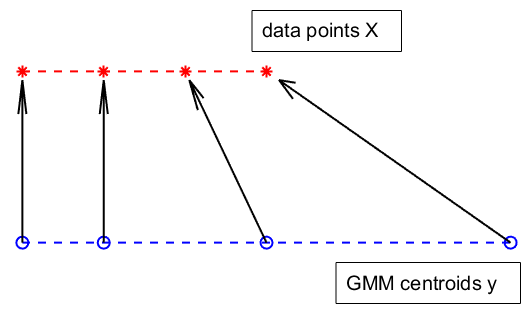}
  \caption{}
  \label{fig:sub-first}
\end{subfigure}
\begin{subfigure}{.5\textwidth}
  \centering
  \includegraphics[width=.9\linewidth]{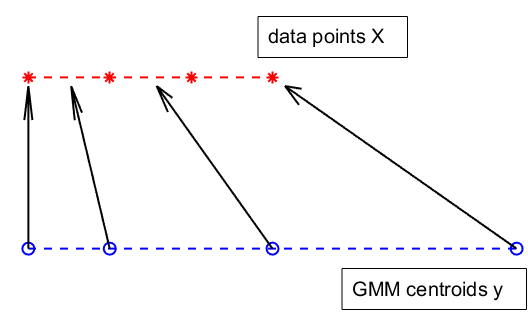}
  \caption{}
\end{subfigure}
\caption{An 1D example under uniform deformation: (a) Results based on the non-rigid point registration of CPD algorithm,
 and(b) The true underlying deformation of scanned data.}
\label{fig:1d example}
\end{figure}

Third, the objectives of CPD algorithm are to obtain correspondence between two point sets and generate the deformation function $\chi$. On the other hand, the objective of material deformation finding algorithm is only to find the  deformation mapping $\chi$.
One example of this difference is shown in Fig. \ref{fig:1d example}, which is a 1d example of a straight line.
The real deformation applied to the original red line is uniform stretching and translation to the bottom. The transformed line is the blue line. However, the locations of the two middle  points in the blue line are chosen arbitrarily which means that the two middle points in the blue line  are not corresponding to the two middle points in the red line.
The registration result of CPD algorithm is shown in Fig. \ref{fig:1d example}(a) where all the points are forced to match. The optimal result of material deformation finding algorithm that we wish to obtain  is shown in Fig. \ref{fig:1d example}(b) where not all GMM centroid $\mathbf{y}$ converge to a data point in $\mathbf{X}$. But the deformation mapping function  generated by the movements of GMM centroid $\mathbf{y}$ is the real $\chi$.

A summary of the differences is shown in the Table. \ref{table:compare between CPD}.

\begin{table}[H]
\caption {Comparison between the point set registration methods and the material deformation finding algorithm.  }
\label{table:compare between CPD}
\begin{center}
    \begin{tabular}{p{6cm} | p{6cm}}

    \toprule
     \textbf{Point set registration methods} & \textbf{Material deformation finding method} \\ \hline
     One to one correspondence for most feature points; & No one to one correspondence for most feature points. \\ \midrule
     Both data sets may contain noises; & Only the scan data contain noises or outliers. \\ \midrule
    Only need feature points on the boundary of the body to define its shape;  & All feature points are inside the domain of the body. \\
    \midrule
    the objectives are to find correspondence and deformation mapping;  & the objective is only to find the deformation mapping.\\
    \bottomrule
    \end{tabular}
\end{center}
\end{table}

\subsection{Motivation}
Based on the comparisons on different aspects of the two methods listed in Table. \ref{table:compare between CPD},
we conclude that the CPD algorithm may not be suitable to solve the material deformation finding problem.
Therefore, a new method need to be proposed.
As discussed above, the original CPD algorithm assumes equal isotropic covariances
and equal membership probabilities for all GMM component.
This assumptions fit well for the classical computer vision problems,
where two point sets are corresponding features extracted from either 2D images or 3D models.
In most cases, there is no underlying mesh where those feature points are selected from.
In our application to 3D printing thermal displacement compensation problem, however,
those two data points $\mathbf{X}_{N \times D}, \mathbf{y}_{m \times D}$ are
vertices of the triangle mesh defined in the STL files.
In fact, the mesh requirement is also not necessary, and
the data points $\mathbf{X}_{N \times D}, \mathbf{y}_{m \times D}$ can serve as
a meshfree representations of the given 3D printed models in different geometric
configurations, i.e. spatial configuration as well as referential configuration.

In meshfree particle method \cite{li2002meshfree, liu2003smoothed},
the geometric state of a continuum body is generally represented by a finite number of discrete particles.
and the motion of the continuum body can be tracked by the motion of these particles.
In the continuum problem domain, each particle
represents a given fraction of the domain, by attaching a small volume of the domain,
and  the mass of each particle is proportional to the volume or area that each particle represents.

\begin{table}[h]
\caption{Pseudo-code for material deformation finding algorithm.}
\label{tbl1}
\begin{tabularx}{\linewidth}{>{\setlength\hsize{1.4\hsize}\setlength\linewidth{\hsize}}X}
\toprule
\begin{itemize}
\item{Initialization: $\mathbf{V}=0, \sigma^{2}=\frac{1}{D N m} \sum_{j=1}^m\sum_{I=1}^{N}\left\|\mathbf{y}_{j}-\mathbf{X}_{I}\right\|^{2}$}
\item{Initialize model parameters: $w, \beta, \lambda, \tilde{W}_m$}
\item{calculate $\mathbf{G}: g_{IJ}= \exp \Bigl( -\frac{1}{2 \beta^{2}}
\left\|\mathbf{y}_{I}-\mathbf{y}_{J}\right\|^{2} \Bigr)$}
\item{ Repeat the following E-step and M-step until convergence is reached:}
    \begin{itemize}
    \item{E-step: Compute $\mathbf{P}$ and $\tilde{\mathbf{P}}$}:
    \begin{itemize}
\item
\[
p_{j I}={
\tilde{w}_m^{1-D/2} \cdot \exp \left(
-\frac{1}{2 \tilde{\sigma}_m^2}
\left\| {\bf X}_I
-{\bf y}_j + G(I, \cdot) \mathbf{V}
\right \|^2 \right )
\over
\displaystyle
\sum_{k=1}^{m}
\tilde{w}_m^{1-D/2} \cdot \exp  \left(
-\frac{1}{2 \tilde{\sigma}_m^2}
\left\|{\bf X}_{I}-  {\bf y}_k+G (K, \cdot) \mathbf{V}
\right \|^2
+\frac{w}{1-w} \frac{( 2 \pi \sigma^{2} )^{D/2} m}{N}
\right )
}
\]
\item $\tilde{p}_{j I} =\frac{p_{j I}}{\tilde{w}_m}
$
\end{itemize}
    \item{$\mathrm{M}$ -step:}
    \begin{itemize}
\item{Solve  ${\bf V}$ based on $\left(\mathbf{G}+\lambda \sigma^{2} d(\tilde{\mathbf{P}} \mathbf{1})^{-1}\right) \mathbf{V}=d(\tilde{\mathbf{P}} \mathbf{1})^{-1} \tilde{\mathbf{P}} \mathbf{y}-\mathbf{X}$ }
\item{ $N_{\tilde{\mathbf{P}}}= \mathbf{1}^{T} \tilde{\mathbf{P}} \mathbf{1}, \mathbf{T}=\mathbf{X}+\mathbf{G} \mathbf{V}$ }

\item{$\sigma^{2}= \frac{1}{N_{P} D}\left(\operatorname{tr}\left(\mathbf{y}^{T} \mathrm{d}\left(\tilde{\mathbf{P}}^{T} \mathbf{1}\right) \mathbf{y}\right)-2 \operatorname{tr}\left((\tilde{\mathbf{P}} \mathbf{y})^{T} \mathbf{T}\right)+\right.
\left.\operatorname{tr}\left(\mathbf{T}^{T} \mathrm{d}(\tilde{\mathbf{P}} \mathbf{1}) \mathbf{T}\right)\right)$
}
      \end{itemize}
    \end{itemize}
\end{itemize} \\
\bottomrule
\end{tabularx}
\end{table}

\subsection{Material deformation finding algorithm}
Inspired by meshfree particle methods e.g. \cite{li2002meshfree},
we assume that the membership probabilities for all GMM components
and their corresponding covariances are proportional to the area that each scan data point assigned based on the underlying mesh.
We can then define the weight  $\tilde{\mathbf{w}}$ for each membership
probability in $\mathbf{y}$ and variance $\sigma^2$ as
\begin{equation}
    \tilde{\mathbf{W}}=\left(\tilde{\mathbf{w}}_{1}, \ldots, \tilde{\mathbf{w}}_{m}\right)^{T} ~~
where ~~ \tilde{\mathbf{w}}_i =
c \cdot area(\mathbf{w}_i)
\end{equation}
where $c$ is a constant. To computer the weight vector  $\tilde{\mathbf{W}}$ and determine the value of $c$, we add one constraint to it: $mean(\tilde{\mathbf{W}}) = 1$, which is
\begin{equation}
    \frac{1}{m}\sum_{j=1}^{m}\tilde{\mathbf{w}}_j = 1
\end{equation}
Therefore, the $j$-th component in GMM
$p(\mathbf{X}_I | \mathbf{y}_j)$ and membership probability $p(j)$ should be modified to
\begin{equation}
p(\mathbf{X}_I | \mathbf{y}_j)=\frac{1}{\left(2 \pi \tilde{\sigma}_j^{2}\right)^{D / 2}} \exp ^{-\frac{\left\|\mathbf{X}_I-\mathbf{y}_{j}\right\| 2}{2 \tilde{\sigma}_j^{2}}} , ~ p(j)=\frac{\tilde{\mathbf{w}}_j}{m}
\label{eq:modify1}
\end{equation}
where $\tilde{\sigma}_j$ is defined as $\tilde{\sigma}^{2}_j =  \tilde{\mathbf{w}}_j  \cdot \sigma^{2}$.

Then the probability density function can be modified as
\begin{equation}
p(\mathbf{X}_I)=w \frac{1}{N}+(1-w) \sum_{j=1}^{m} \frac{\tilde{\mathbf{w}}_j}{m} p(\mathbf{X}_I | \mathbf{y}_j)
\label{eq:GMM_new}
\end{equation}

After recasting Eq. (\ref{eq:GMM}) into Eq. (\ref{eq:GMM_new}), the material deformation finding problem could be represented as a MLE and solved by EM algorithm as in original CPD algorithm.
However, there is still one question need to be answered:
Since there is no corresponding pairs $(\mathbf{X}_I,\mathbf{x}_j)$ between two point sets, why we continue using MLE to "match " two sets of points  and what is the result of the "matching"?

\begin{figure}[ht]
\begin{subfigure}{.5\textwidth}
  \centering
  \includegraphics[width=.9\linewidth]{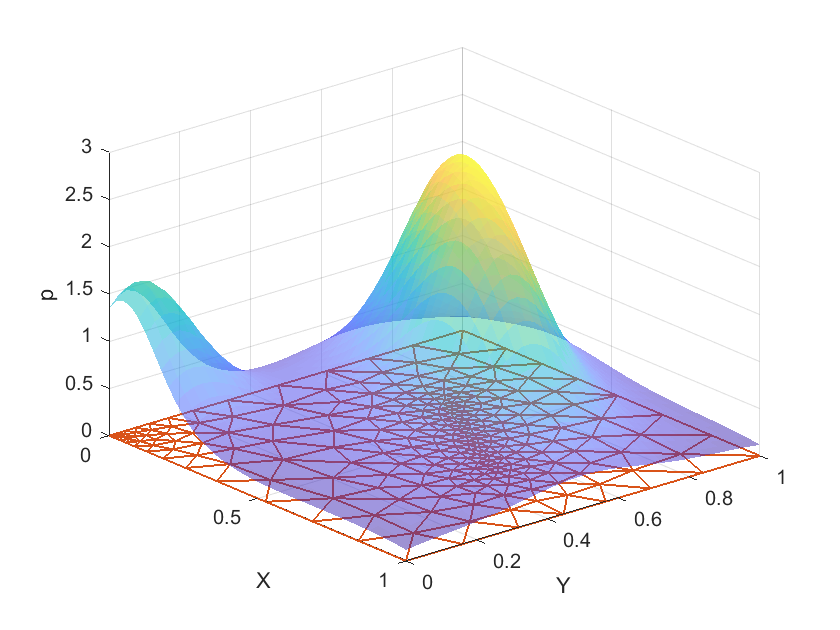}
  \caption{}
  \label{fig:sub-first}
\end{subfigure}
\begin{subfigure}{.5\textwidth}
  \centering
  \includegraphics[width=.9\linewidth]{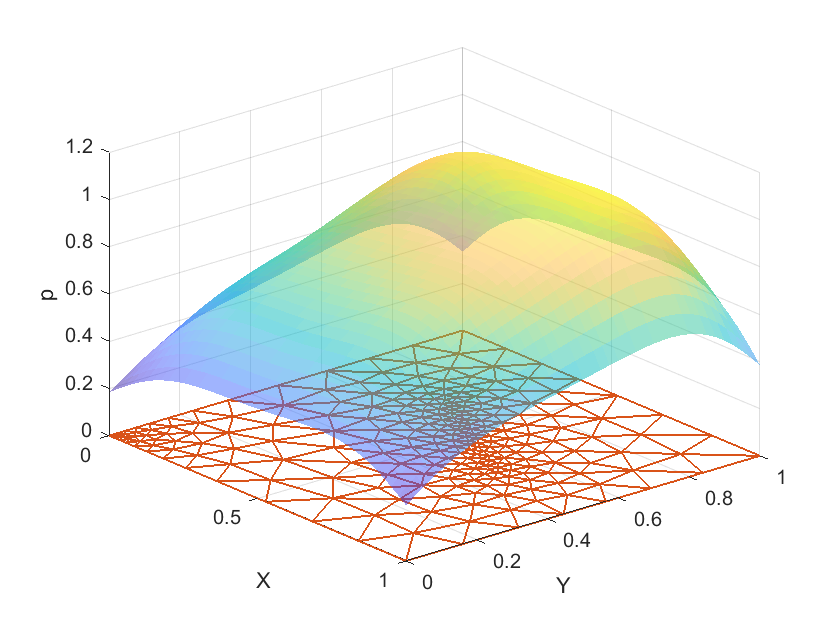}
  \caption{}
\end{subfigure}
\medskip
\begin{subfigure}{0.5\textwidth}
\centering
  \includegraphics[width=.9\linewidth]{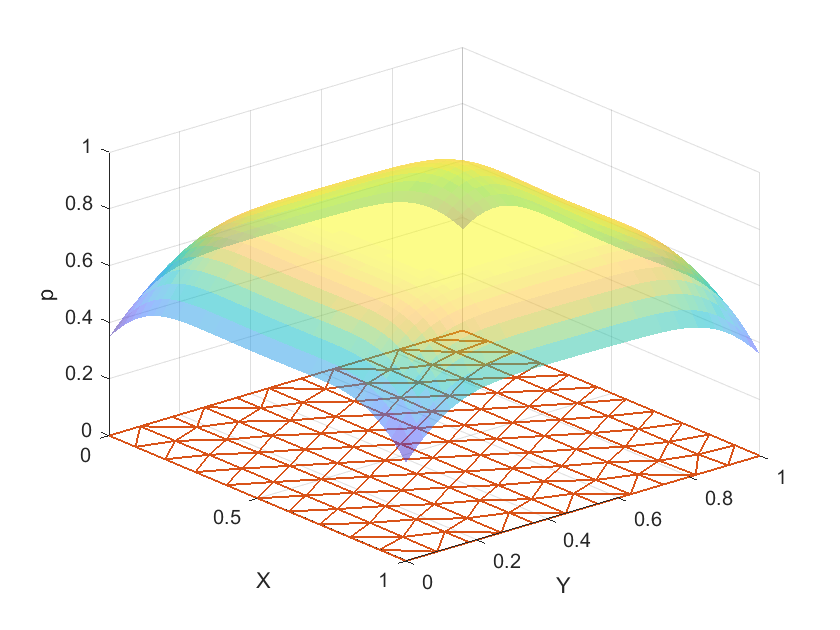}
\caption{}
\end{subfigure}\hfil 
\caption{GMM density function: (a) non-uniform mesh with equal membership probability. (b)non-uniform mesh with specific membership probability for each member. (c)uniform mesh with equal membership probability.}
\label{fig:density}
\end{figure}

The underlying assumption of this specific membership probability is that feature points are mesh-free representation of the model and the shape of the model should be independent of how to discretize it.
Therefore, ideally, the GMM density function generated by any arbitrary mesh should be identical to the GMM density function generated by the uniform mesh.
And the registration between a uniform mesh and an arbitrary mesh is conceptually identical to a registration problem between two same meshes which can be solved by CPD algorithm. For example, the GMM density function of a 2D model generated by different setting is shown in Fig. \ref{fig:density}.
The Fig. \ref{fig:density} illustrates that the GMM density functions generated by uniform mesh and non-uniform mesh with specific membership probability for each member follow the same trend in the unit domain. However, GMM density function of  the non-uniform mesh with equal membership probability has two sharp modes where mesh are dense which make it a poor representation of the true underlying density function.
A numerical experiment of this model will be further studied in section 5.1.

By modifying the EM algorithm,
we can write the objective function $\mathcal{L}_Q$
in Eq.(\ref{eq:Q function}) as,
\begin{equation}
\mathcal{L}_Q\left(\theta, \sigma^{2}\right)=\frac{1}{2 \sigma^{2}} \sum_{n=1}^{N} \sum_{j=1}^{m} p^{\mathrm{old}}\left(\mathbf{y}_{j} | \mathbf{X}_{n}\right)
\frac{\left\|\mathbf{X}_{n}-\chi^{-1}\left(\mathbf{y}_{j}, \theta\right)\right\|^{2}}{\tilde{w}_j}+
\frac{N_{p} D}{2} \log \sigma^{2}
\label{eq:Q function3}
\end{equation}
where
$p^{\text {old }}$
is a conditional probability of each designed mesh nodal point
to each member of GMM density function by using model parameters from the late iteration step,
\begin{equation}
p^{old} \left(\mathbf{y}_{j} | \mathbf{X}_{n} \right)= \displaystyle
\displaystyle
{
\exp \Bigl(
-\frac{1}{2}
\left\|\frac{{\bf X}_{n}-\chi^{-1}
\left( {\bf y}_{j}, \theta^{old} \right)}
{\tilde{\sigma}^{old}}
\right\|^{2}
\Bigr)
\over
\displaystyle \sum_{j=1}^{m} \exp \left(
{-\frac{1}{2}
\left\|
\frac{ {\bf X}_{n}-\chi^{-1}\left({\bf y}_{j}, \theta^{old} \right)}
{\tilde{\sigma}^{old}}
\right\|^{2}+c}
\right)
}
\end{equation}
where $c=\left(2 \pi \sigma^{2}\right)^{D / 2} \frac{w}{1-w} \frac{m}{N} .$
In order to coerce GMM centroid to move in a consistent manner, we use a
regularization term that is similar in CPD algorithm as
\begin{equation}
E(\mathbf{V})=\operatorname{Tr}\left(\mathbf{V}^{T} \mathbf{G} \mathbf{V}\right)
\end{equation}
where $\mathbf{G}_{m \times m}$ again is the Gaussian kernel matrix and $\mathbf{V}_{m \times D}=(\mathbf{v}_1,...,\mathbf{v}_m)^T$ is the coefficients matrix that may be interpreted as
a pseudo-velocity, which determine the displacements.
With this regularization term, the objective function in Eq.(\ref{eq:Q function}) can be rewritten as
\begin{equation}
\mathcal{L}_Q\left(\theta, \sigma^{2}\right)=\frac{1}{2 \sigma^{2}} \sum_{n=1}^{N} \sum_{j=1}^{m} p^{\mathrm{old}}\left(\mathbf{y}_{j} | {\bf X}_{n}\right)
\frac{\left\|{\bf X}_{n}-\chi^{-1}\left( {\bf y}_{j}, \theta\right)\right\|^{2}}{\tilde{w}_j}+
\frac{N_{p} D}{2} \log \sigma^{2}
+ \frac{\alpha}{2}
\operatorname{Tr}\left(\mathbf{V}^{T} \mathbf{G} \mathbf{V}\right)
\label{eq:Q function4}
\end{equation}
The transformation of GMM centroid can be defined as
a generalized radial basis function which can be derive form Eq. (\ref{eq:regularization})
by using the representation theorem in the reproducing kernel Hilbert space (RKHS) (see \cite{scholkopf2001generalized}),
\begin{equation}
\chi(\mathbf{y}, \mathbf{V})=\mathbf{y}+\mathbf{G} \mathbf{V}
\end{equation}
Thus, the objective function in Eq.(\ref{eq:Q function4})
can be written in matrix form as,
\begin{equation}
\begin{aligned}
\mathcal{L}_Q \left(\mathbf{V}, \sigma^{2}\right)=& \frac{1}{2 \sigma^{2}}\left\{\operatorname{Tr}\left(\mathbf{y}^{T} d\left(\tilde{\mathbf{P}}^{T} \mathbf{1}\right) \mathbf{y}\right)
- 2 \operatorname{Tr}\left(\mathbf{X}^{T} \tilde{\mathbf{P}} \mathbf{y}\right)-2 \operatorname{Tr}\left(\mathbf{V}^{T} \mathbf{G} \tilde{\mathbf{P}} \mathbf{y}\right)+\operatorname{Tr}\left(\mathbf{X}^{T} d(\tilde{\mathbf{P}} \mathbf{1}) \mathbf{X}\right)\right\} \\
+& 2 \operatorname{Tr}\left(\mathbf{V}^{T} \mathbf{G} d(\tilde{\mathbf{P}} \mathbf{1}) \mathbf{X}\right)+\operatorname{Tr}\left(\mathbf{V}^{T} \mathbf{G} d(\tilde{\mathbf{P}} \mathbf{1}) \mathbf{G} \mathbf{V}\right)
+ \frac{N_{p} D}{2} \ln \left(\sigma^{2}\right)+\frac{\alpha}{2} \operatorname{Tr}\left(\mathbf{V}^{T} \mathbf{G} \mathbf{V}\right)
\label{eq:Q function5}
\end{aligned}
\end{equation}
where $\mathbf{P}$ is a $m \times N$ matrix with element $p^{o l d}\left(\mathbf{y}_{j} | \mathbf{X}_{n}\right)$,
$d(\mathbf{V})$ is a diagonal matrix with elements
on the diagonal equal the value of a vector $\mathbf{V}$, $\mathbf{1}$
is a column vector where all the elements equal one.

$\mathcal{L}_Q\left(\mathbf{V}, \sigma^{2}\right)$ is a quadratic function of coefficient matrix $\mathbf{V}$.
With the aim of updating $\mathbf{V}$ that minimizes the objective function in Eq.(\ref{eq:Q function5}),
we can set the derivative of function $\mathcal{L}_Q\left(\mathbf{V}, \sigma^{2}\right)$ to zero, which is
\begin{equation}
\frac{\partial \mathcal{L}_Q\left(\mathbf{V}, \sigma^{2}\right)}{\partial \mathbf{V}}=-\mathbf{G} \tilde{\mathbf{P}} \mathbf{y}+\mathbf{G} d(\tilde{\mathbf{P}} \mathbf{1}) \mathbf{X}+\mathbf{G} d(\tilde{\mathbf{P}} \mathbf{1}) \mathbf{G} \mathbf{V}
+\sigma^{2} \alpha \mathbf{G} \mathbf{V} =0~.
\label{eq:Eq-W}
\end{equation}
Then, we can find the value of coefficients $\mathbf{V}$ by solving the equation below,
\begin{equation}
\left[d(\tilde{\mathbf{P}} \mathbf{1}) \mathbf{G}+\sigma^{2} \alpha \mathbf{I}\right] \mathbf{V}
=\tilde{\mathbf{P}} \mathbf{X}-d(\tilde{\mathbf{P}} \mathbf{1}) \mathbf{X}
\end{equation}
Similarly, we can update $\sigma^{2}$ as,
\begin{equation}
\sigma^{2}= \frac{1}{N_{P} D}\left(\operatorname{tr}\left(\mathbf{y}^{T} \mathrm{d}\left(\tilde{\mathbf{P}}^{T} \mathbf{1}\right) \mathbf{y}\right)-2 \operatorname{tr}\left((\tilde{\mathbf{P}} \mathbf{X})^{T} \mathbf{T}\right)+\right.
\left.\operatorname{tr}\left(\mathbf{T}^{T} \mathrm{d}(\tilde{\mathbf{P}} \mathbf{1}) \mathbf{T}\right)\right)~.
\end{equation}
To implement the above computational formulation,
a pseudo-code of the material deformation finding algorithm is given in Table \ref{tbl1}.

\begin{remark}
In the proposed material deformation finding algorithm,
there are three free parameters: $w,~ \lambda,~ \alpha$.
Parameter $w$ implies our prior belief of the level noise in the data set;
And both $\lambda$ and $\alpha$ determine the level of motion coherence that one wants to
enforce.
We can tune all three free parameters until finding the best matching result.
\end{remark}

\section{Experiments and results}
In this Section, we shall present two examples to demonstrate the accuracy, robustness,
and improvement of the material deformation finding algorithm over the original CPD algorithm.
The first example is a synthetic 2D problem in which the designed
product is a two-dimensional unit square with
a uniform triangle mesh and the scanned material data is a non-uniform or locally refined mesh on the deformed unit square.
The material registration results
are compared with the original CPD algorithm under same values of algorithm parameters.
The second experiment is a real application of 3D printing example (see Fig. \ref{Example of Distortion}).
The proposed material deformation finding algorithm
 is used to find the residual or permanent deformation of the 3D printed product in Fig. \ref{Example of Distortion}
 due to temperature effect in a additive manufacturing process.
 Based on the predicted residual deformation, one can find
 the thermal compensation or compensation allowance, such that
 a new specimen may be printed on a compensated mold,
 and the final result is compared with the first printed specimen.

\begin{figure}[h]
    \centering 
\begin{subfigure}{0.3\textwidth}
  \includegraphics[width=1.7in]{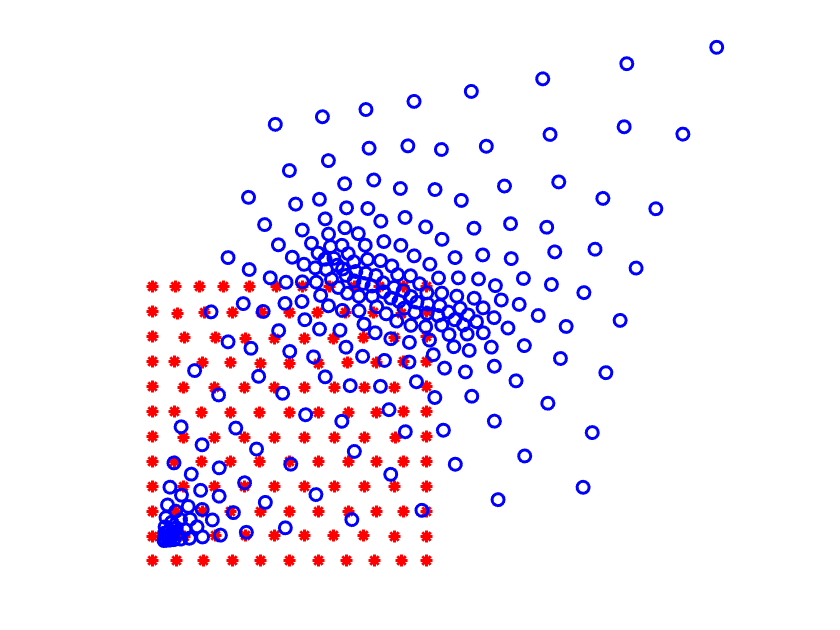}
  \label{fig:1}
\end{subfigure}\hfil 
\begin{subfigure}{0.3\textwidth}
  \includegraphics[width=1.7in]{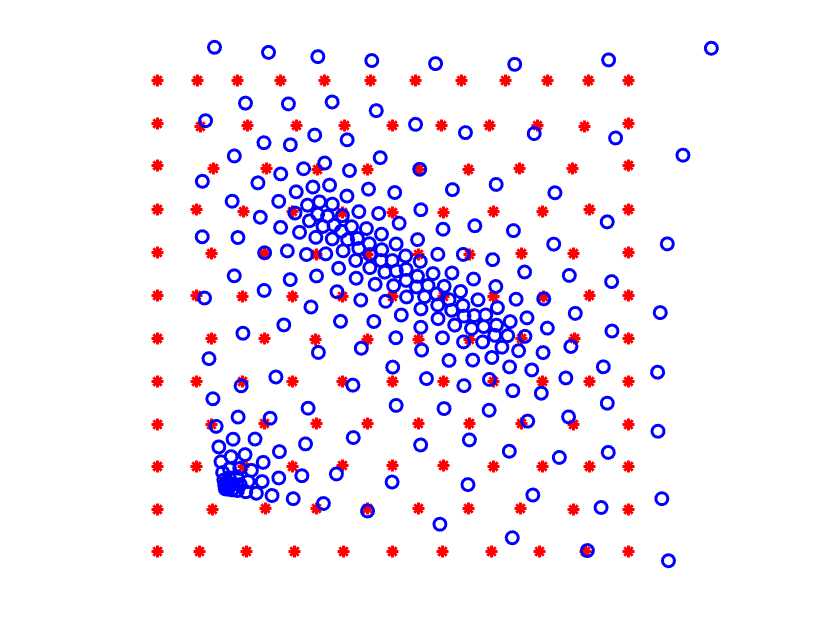}
  \label{fig:2}
\end{subfigure}\hfil 
\begin{subfigure}{0.3\textwidth}
  \includegraphics[width=1.7in]{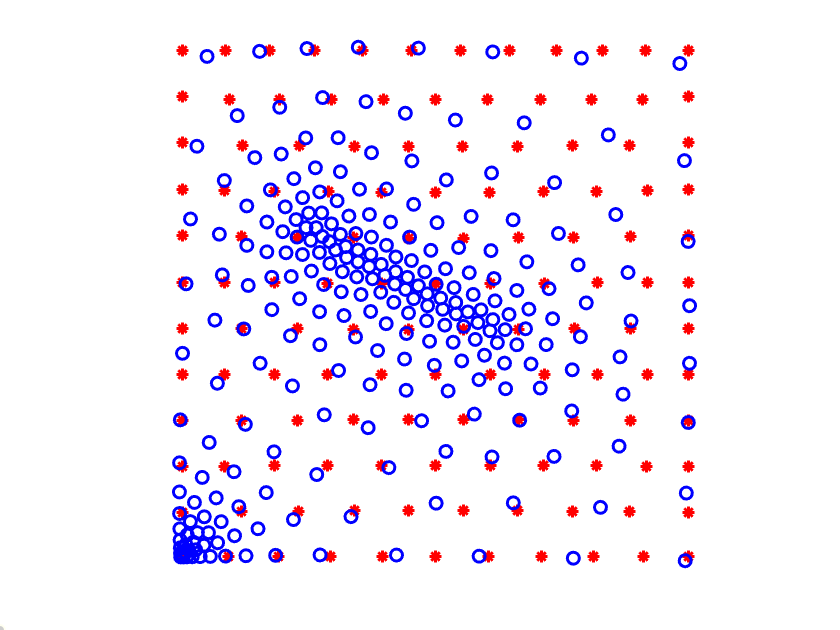}
  \label{fig:3}
\end{subfigure}
\medskip
\begin{subfigure}{0.3\textwidth}
  \includegraphics[width=1.7in]{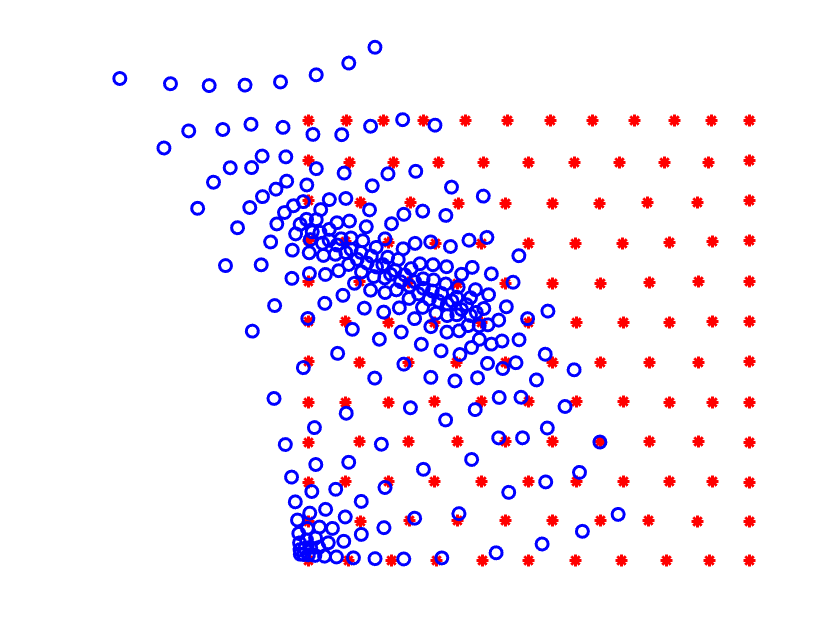}
  \label{fig:4}
\end{subfigure}\hfil 
\begin{subfigure}{0.3\textwidth}
  \includegraphics[width=1.7in]{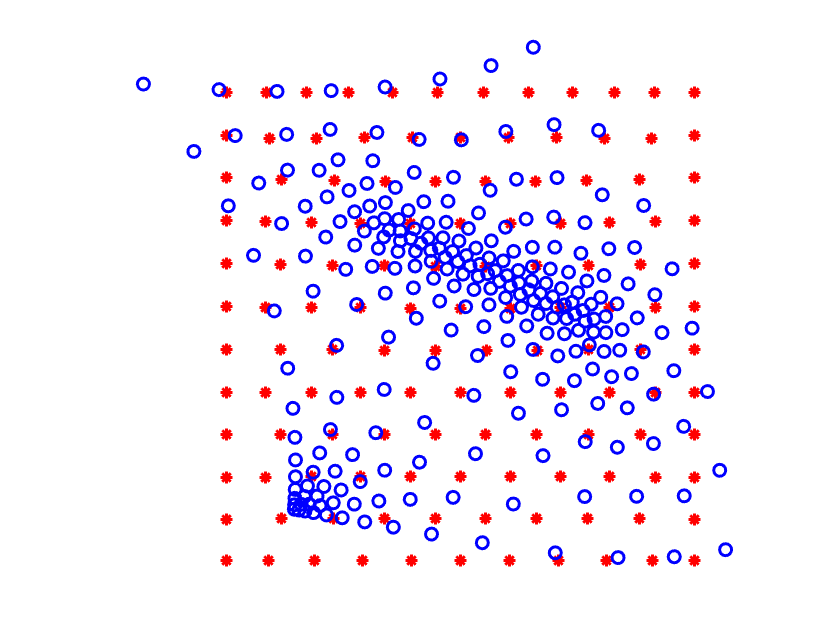}
  \label{fig:5}
\end{subfigure}\hfil 
\begin{subfigure}{0.3\textwidth}
  \includegraphics[width=1.7in]{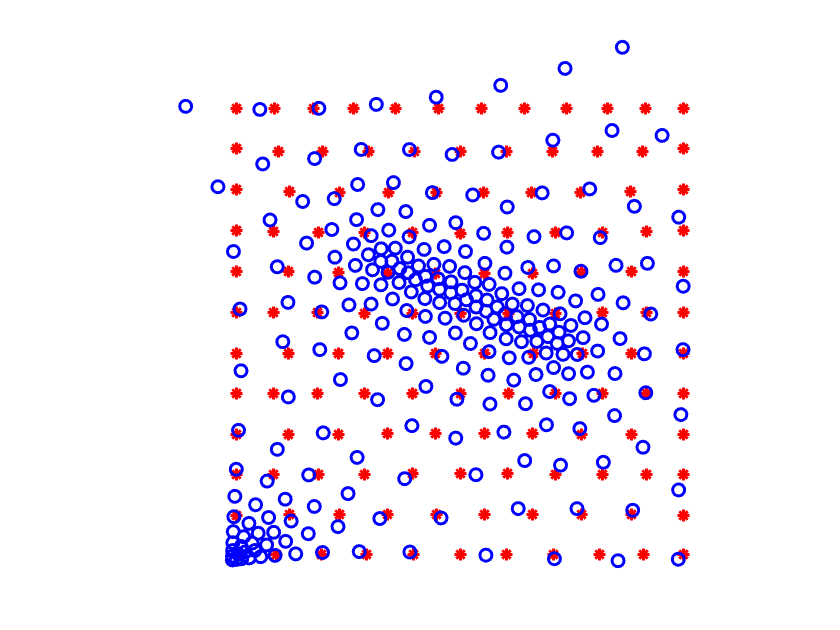}
  \label{fig:6}
\end{subfigure}
\medskip
\begin{subfigure}{0.3\textwidth}
  \includegraphics[width=1.7in]{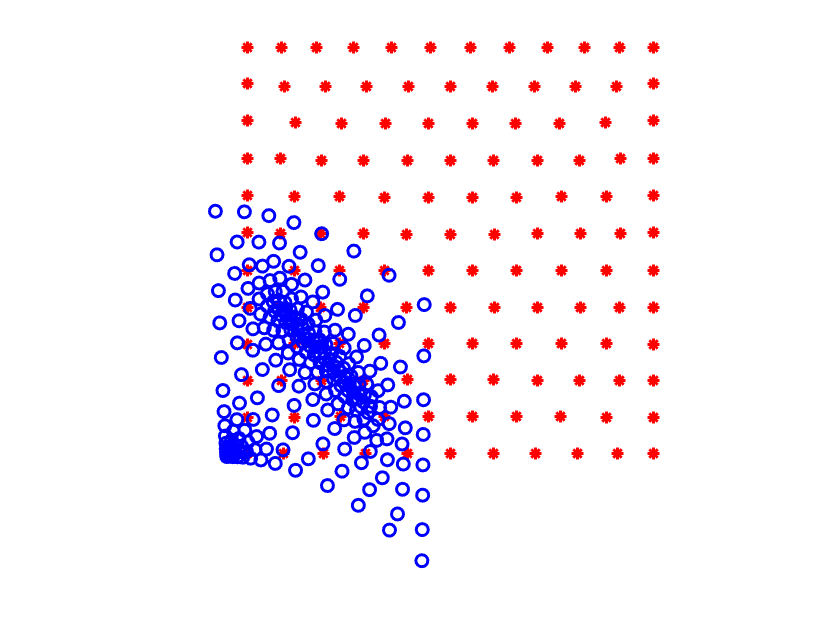}
  \caption{}
  \label{fig:7}
\end{subfigure}\hfil 
\begin{subfigure}{0.3\textwidth}
  \includegraphics[width=1.7in]{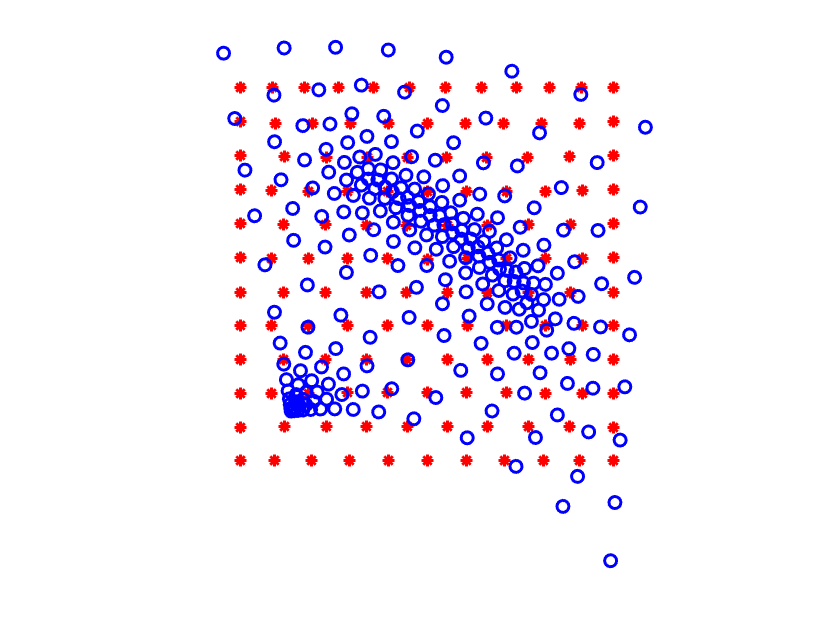}
  \caption{}
  \label{fig:8}
\end{subfigure}\hfil 
\begin{subfigure}{0.3\textwidth}
  \includegraphics[width=1.7in]{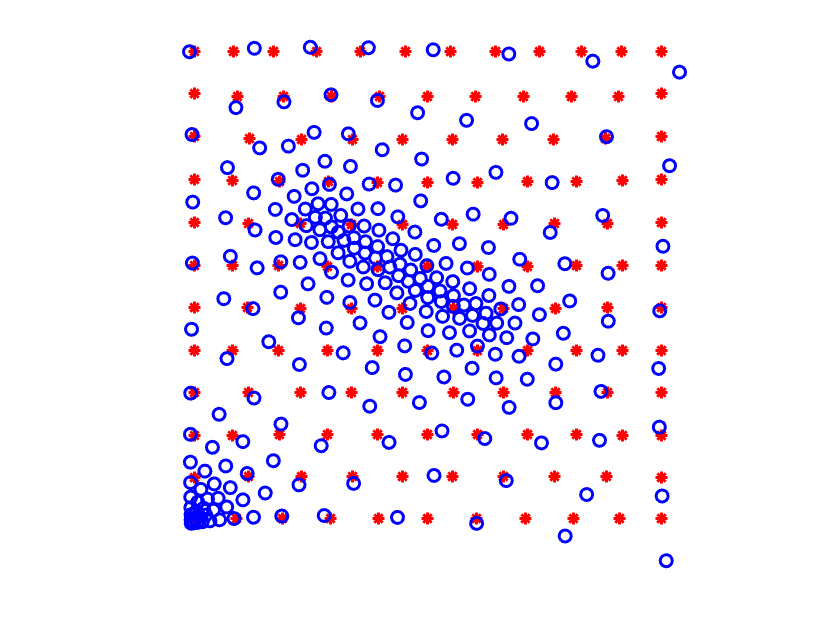}
  \caption{}
  \label{fig:9}
\end{subfigure}
\caption{2D point set registration results of the proposed material deformation finding algorithm and the original CPD algorithm with different nonrigid deformation. (a) three different nonrigid deformation before registration. (b) the registration results of CPD algorithm. (c) the registration results of material  deformation  finding  algorithm.}
\label{fig:square registration}
\end{figure}

\subsection{Registration between uniform mesh and locally refined non-uniform mesh}
As mentioned before,
the main challenge in the determination of the distortion displacement
field for a 3D printed structure component is that
we only know the spatial positions of material points in the current configuration,
but we do not know the corresponding material coordinates in the referential configuration.
In order to find the displacement field, we have to find
the material coordinates of all spatial material points by material registration.
To demonstrate the effectiveness of material registration process based on
the proposed material deformation finding algorithm, in the first example,
we choose a 2D specimen of a unit square as the shape in the referential configuration,
which is first meshed with uniform triangle mesh whose nodal points are serving as
the reference points, or data points.
The uniform mesh in the referential configuration is shown in Fig.\ref{fig:mesh} II(a),
which we also call as the design mesh in the referential configuration.
Then we create a locally refined non-uniform mesh by Delaunay triangulation algorithm
\cite{persson2004simple},
which is still in the referential configuration as shown in Fig.\ref{fig:mesh} I(a).
After that, we apply non-zero strain or non-rigid deformation to the square specimen,
and it will deform into a near rhombic shape as shown Fig.\ref{fig:mesh} I(b).
The spatial positions of the nodal points of the locally refined
non-uniform mesh are assumed as the scanned material data (see Fig.\ref{fig:mesh} I(b)),
which is usually known by scanning the 3D printed product.
In this way, the nodal point positions in the referential configuration i.e.
Fig.\ref{fig:mesh} I (a) are their material coordinates,
which are the exact solution of the material registration of this synthetic problem,
and they are unknown for any practical 3D printing problem.
In this example, we are trying to use the proposed material deformation finding algorithm
to recover the material coordinates for the scanned material points whose
spatial positions as shown in Fig.\ref{fig:mesh} I(b), based on
a set of reference or designed nodal points shown in Fig.\ref{fig:mesh} II(a).
Finally, we can compare the numerical solution with the exact solution, i.e.
Fig.\ref{fig:mesh} III(a).

In actual numerical experiments, three different arbitrary nonuniform deformation are generated by using
polynomial displacement field with random coefficients.
Three different sets of random coefficients are chosen with different levels of non-uniform deformation, i.e.
with non-zero strain distributions. Then we apply these non-uniform motion or deformation to
the unit square plate (see Fig. \ref{fig:mesh} I).
In numerical experiment discussed below, we set the system parameters' values for the material deformation
finding algorithm as $ \beta=3, \alpha =2$, and $w =0$ for all three cases.

\begin{figure}[h]
\centering
\begin{subfigure}{0.3\textwidth}
  \includegraphics[width=\linewidth]{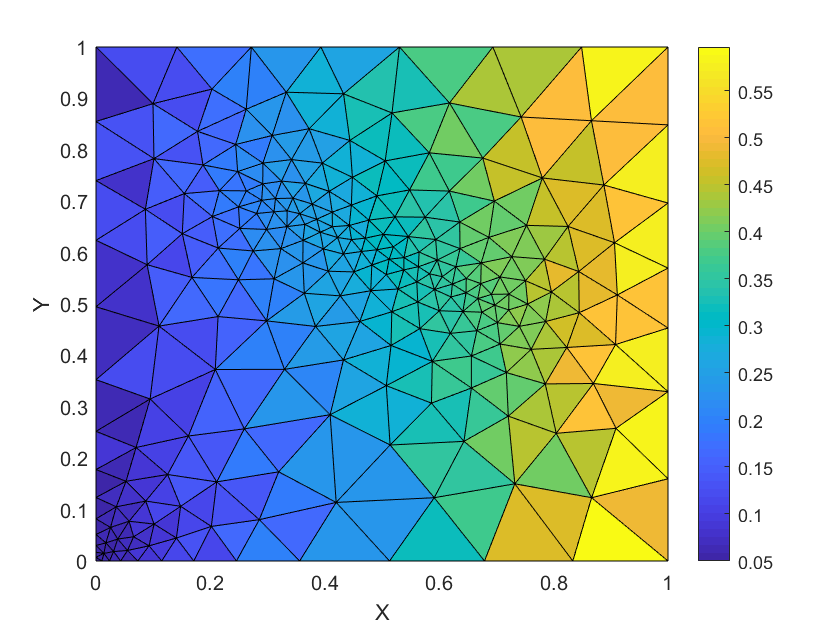}
\caption{applied displacement in x-axis}
  \label{fig:1}
\end{subfigure}\hfil 
\begin{subfigure}{0.3\textwidth}
  \includegraphics[width=\linewidth]{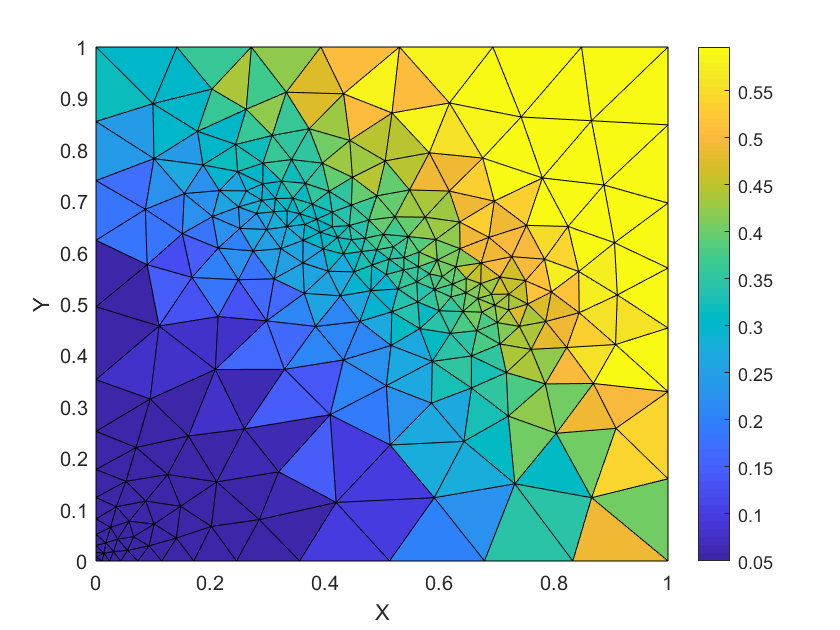}
\caption{displacement field calculated by CPD in x-axis}
  \label{fig:2}
\end{subfigure}\hfil 
\begin{subfigure}{0.3\textwidth}
  \includegraphics[width=\linewidth]{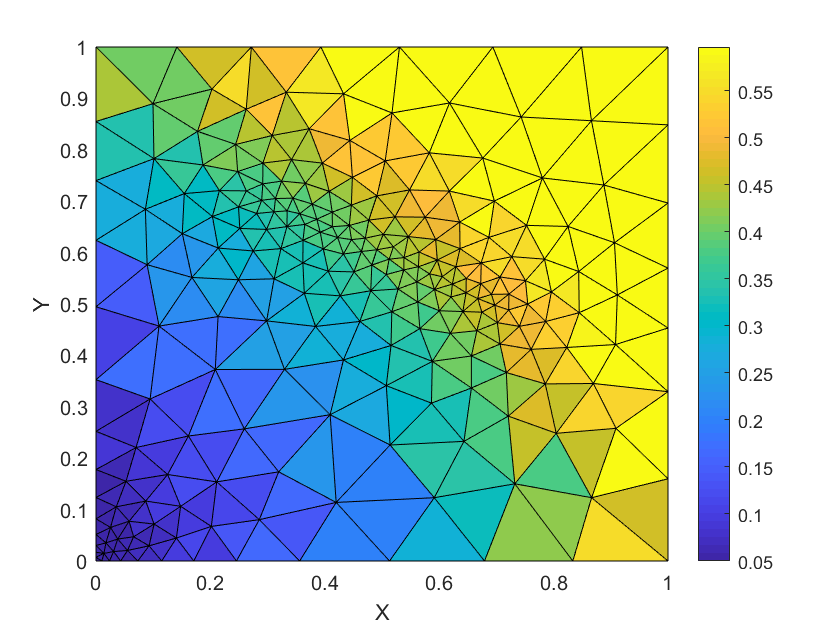}
\caption{displacement field calculated by material deformation finding algorithm in x-axis}
  \label{fig:3}
\end{subfigure}
\medskip
\begin{subfigure}{0.3\textwidth}
  \includegraphics[width=\linewidth]{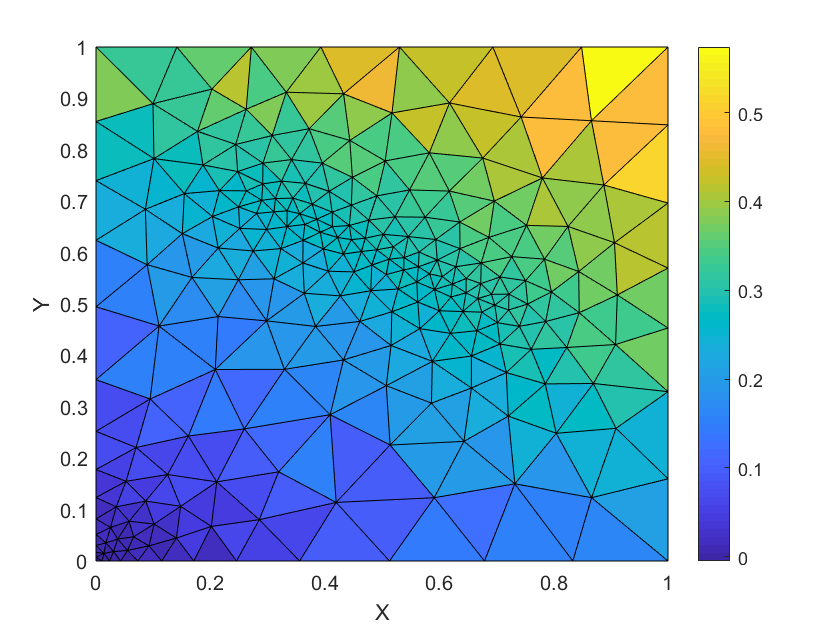}
\caption{applied displacement in y-axis}
  \label{fig:4}
\end{subfigure}\hfil 
\begin{subfigure}{0.3\textwidth}
  \includegraphics[width=\linewidth]{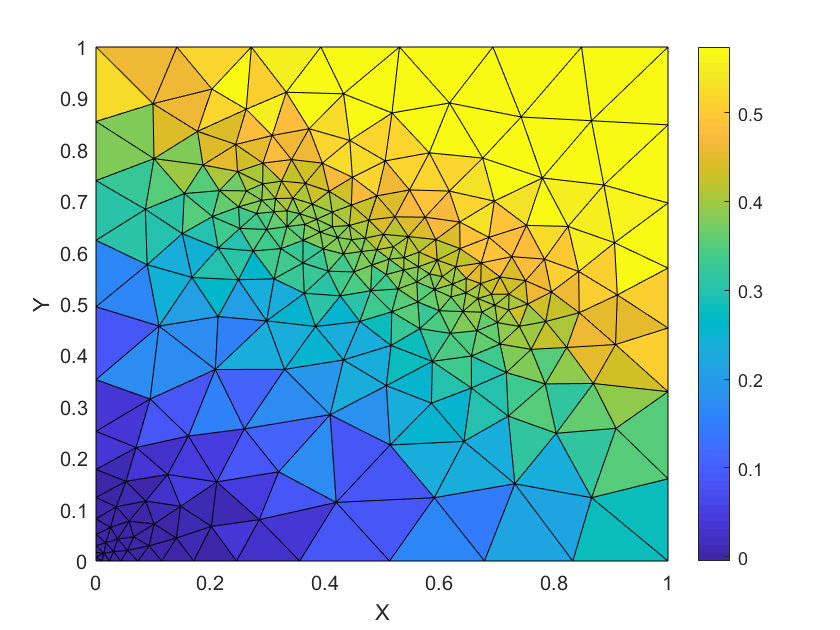}
\caption{displacement field calculated by CPD in y-axis}
  \label{fig:5}
\end{subfigure}\hfil 
\begin{subfigure}{0.3\textwidth}
  \includegraphics[width=\linewidth]{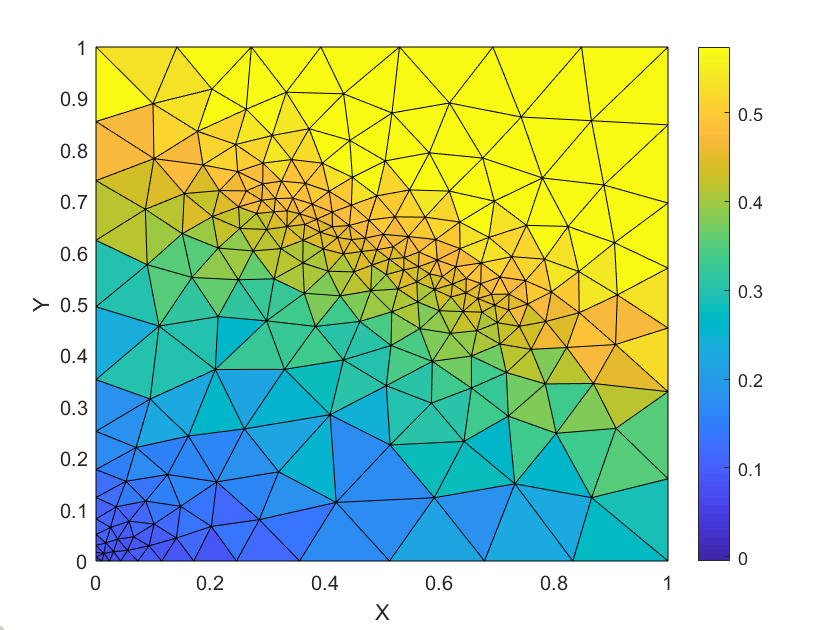}
\caption{displacement field calculated by material deformation finding algorithm in y-axis}
\end{subfigure}
\caption{Displacement field of unit square for case 1.
The first row represents the value of displacement in x-axis.
The second row shows the value of displacement in y-axis.
The right column plots the assigned displacement field in both direction.
The displacement field calculated by original CPD algorithm is shown in the middle column.
The displacement field calculated by material  deformation  finding  algorithm is shown in the right column.}
\label{fig:images}
 \label{fig:Disp-filed-case1}
\end{figure}

Figure \ref{fig:square registration}
shows non-uniform material deformation registration results for three different prescribed
deformation cases.
The left column shows three different experiments the design(red) and scanned(blue) models.
The middle column shows the registration results of CPD algorithm.
The right column represents the registration results of the material deformation finding algorithm.
Then we can interpolate the displacement fields in both $x$ and $y$ directions.
The displacement distributions for case 1 are shown in Fig. \ref{fig:Disp-filed-case1} by color contour plots.
The first row represents the value of displacement in x-axis.
The second row shows the value of displacement in y-axis.
The right column plots the assigned displacement field in both direction.
The displacement field calculated by CPD algorithm is shown in the middle column.
The displacement field calculated by material  deformation  finding  algorithm is shown in the right column.
It can be seen from Fig. \ref{fig:Disp-filed-case1}
that non-uniform deformation for all cases can be well captured
by the proposed material deformation finding algorithm,
especially for the boundary points that
define the shape of the domain, whereas
the original CPD method fails to register two data sets.

\begin{figure}[h]
	\centering
		\includegraphics[width=4.0in]{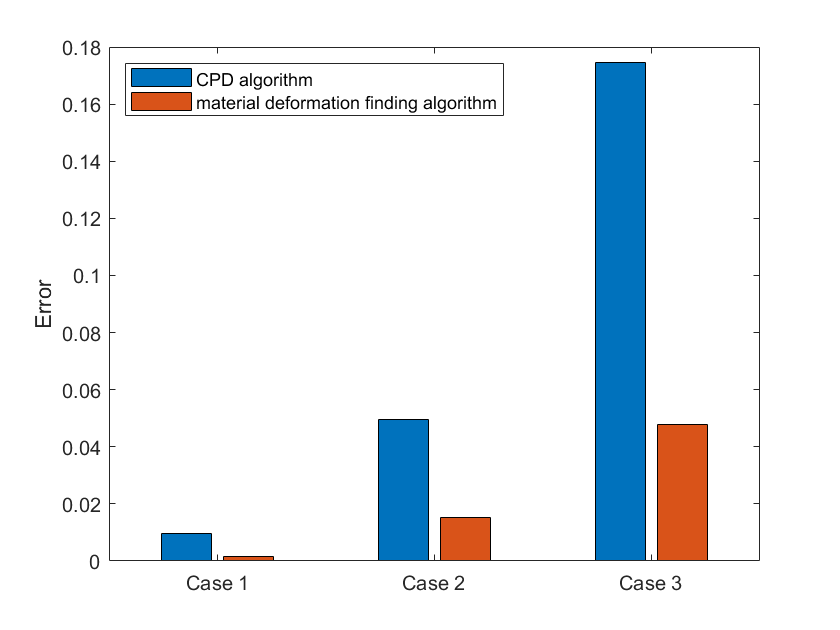}
	\caption{Average registration error for four corner points.}
	\label{FIG:corner error}
\end{figure}

Also, in order to quantitatively compare the results of CPD algorithm and material deformation finding algorithm, the registration errors of fours corner points in all three cases are calculated based on Euclidean distance as illustrated in Fig. \ref{FIG:corner error}.
For each level of non-uniform deformation, our new material deformation finding algorithm always yields a significant better result than the CPD algorithm.

\begin{figure}[h]
	\centering
		\includegraphics[width=4.0in]{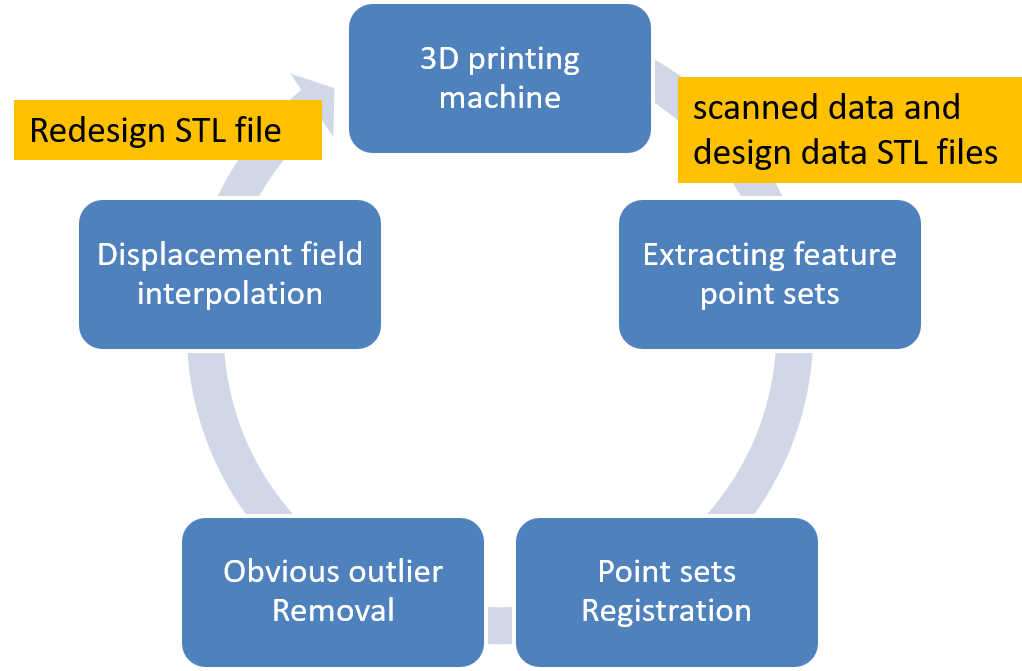}
	\caption{Flowchart of the data pipeline for thermal deformation compensation.}
	\label{FIG:Flowchart}
\end{figure}

\subsection{Additive manufacturing material deformation finding algorithm}

The proposed material deformation finding algorithm is applied to  3D printing thermal distortion problem which is solely based on the scanned data
of material point position of the printed product and the designed data of the product.
The main idea is that: (1) The material displacement field is first obtained by material deformation finding algorithm. (2) The second step is to redesign the shape of product and the related 3D printing file or mold
in order to obtain the final un-distorted printed product.

\begin{figure}[h]
  \begin{subfigure}[b]{0.45\textwidth}
    \includegraphics[width=2.5in]{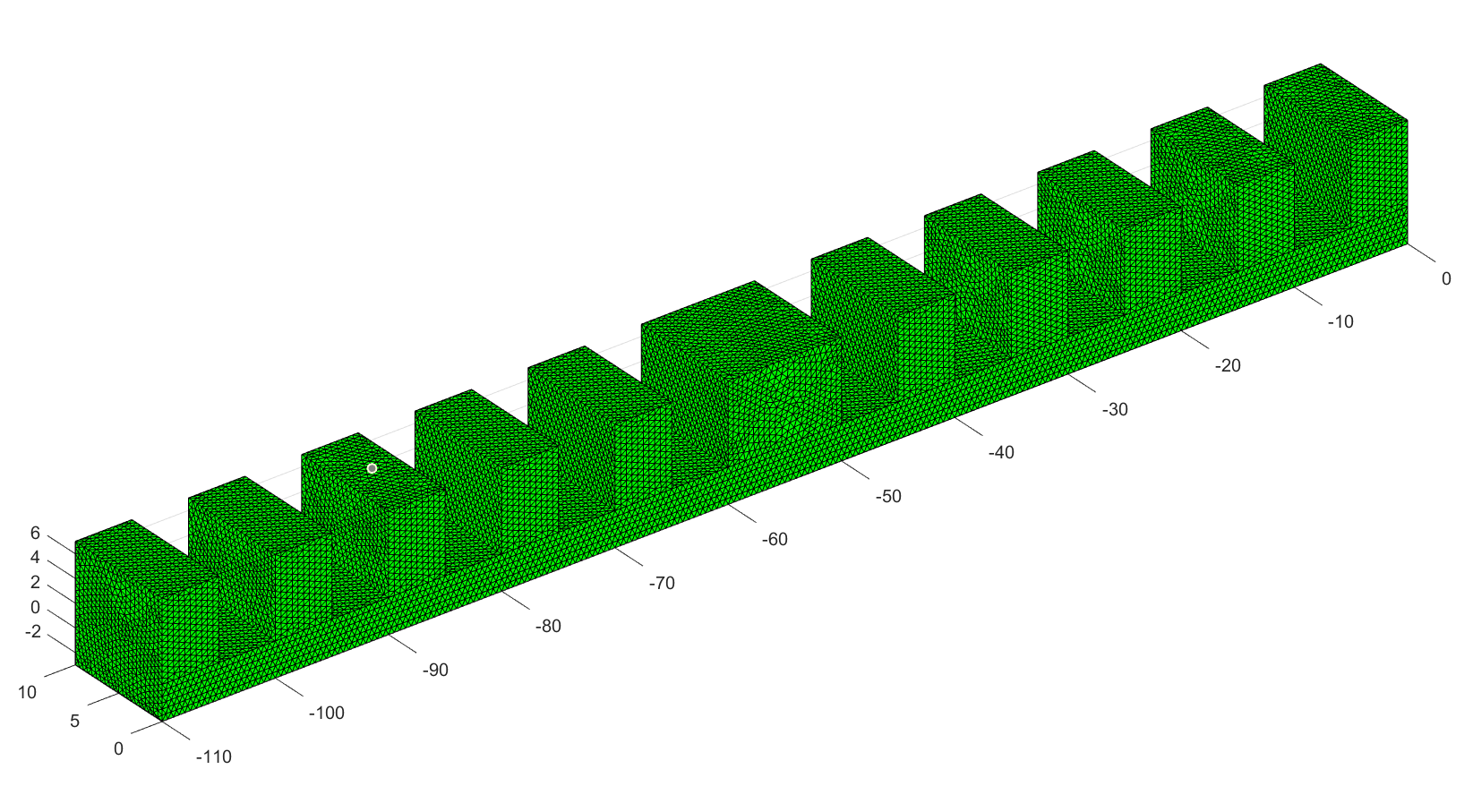}
    \caption{3D view.}
  \end{subfigure}
  \hfill
  \begin{subfigure}[b]{0.48\textwidth}
    \includegraphics[width=2.5in]{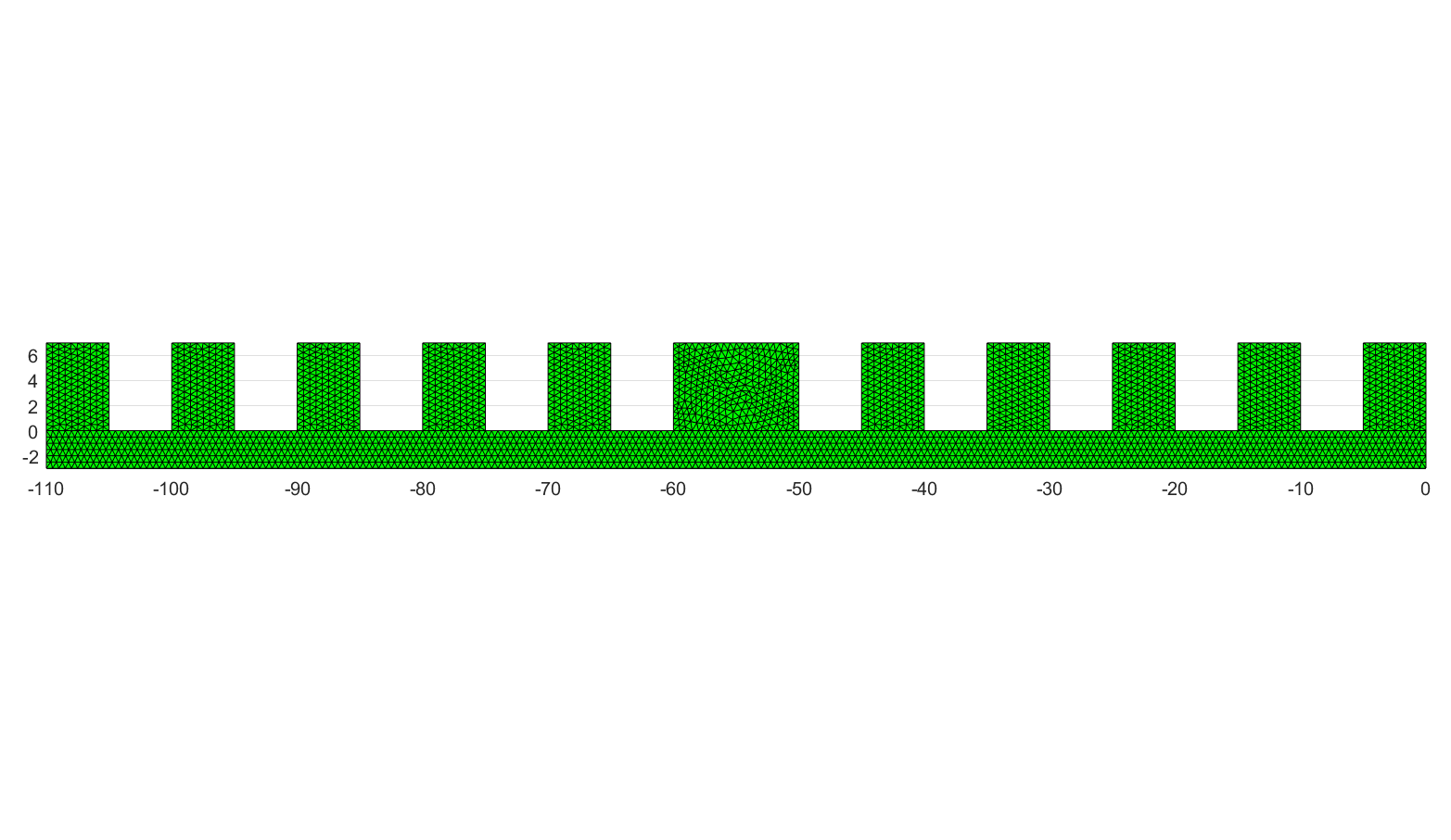}
    \caption{Front view.}
  \end{subfigure}
  \caption{Referential mesh of design model in the initial or material configuration.}
  \label{fig:design_data}
\end{figure}

\begin{figure}[h]
  \begin{subfigure}[b]{0.48\textwidth}
    \includegraphics[width=2.5in]{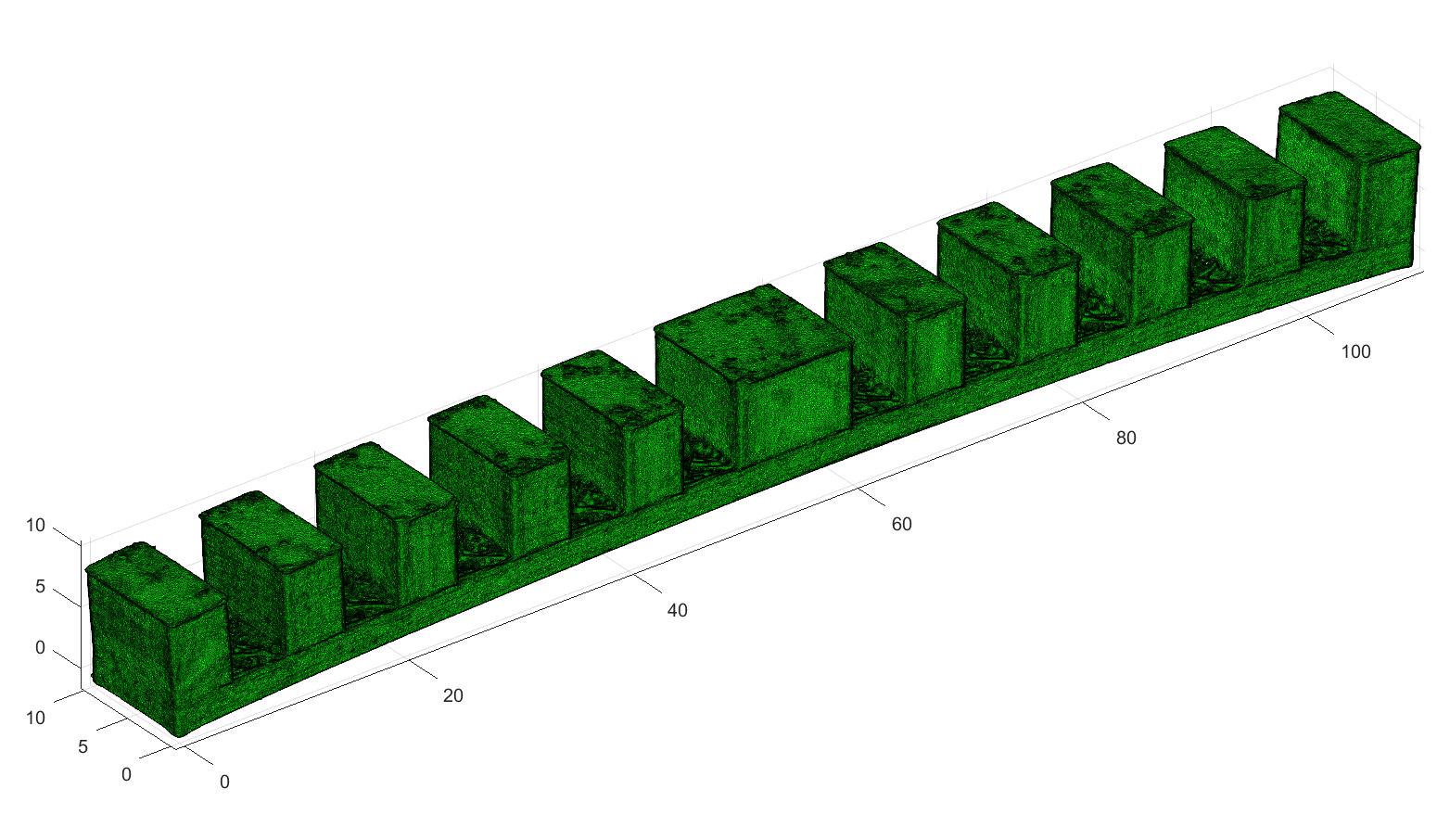}
    \caption{3D view.}
  \end{subfigure}
  \hfill
  \begin{subfigure}[b]{0.48\textwidth}
    \includegraphics[width=2.5in]{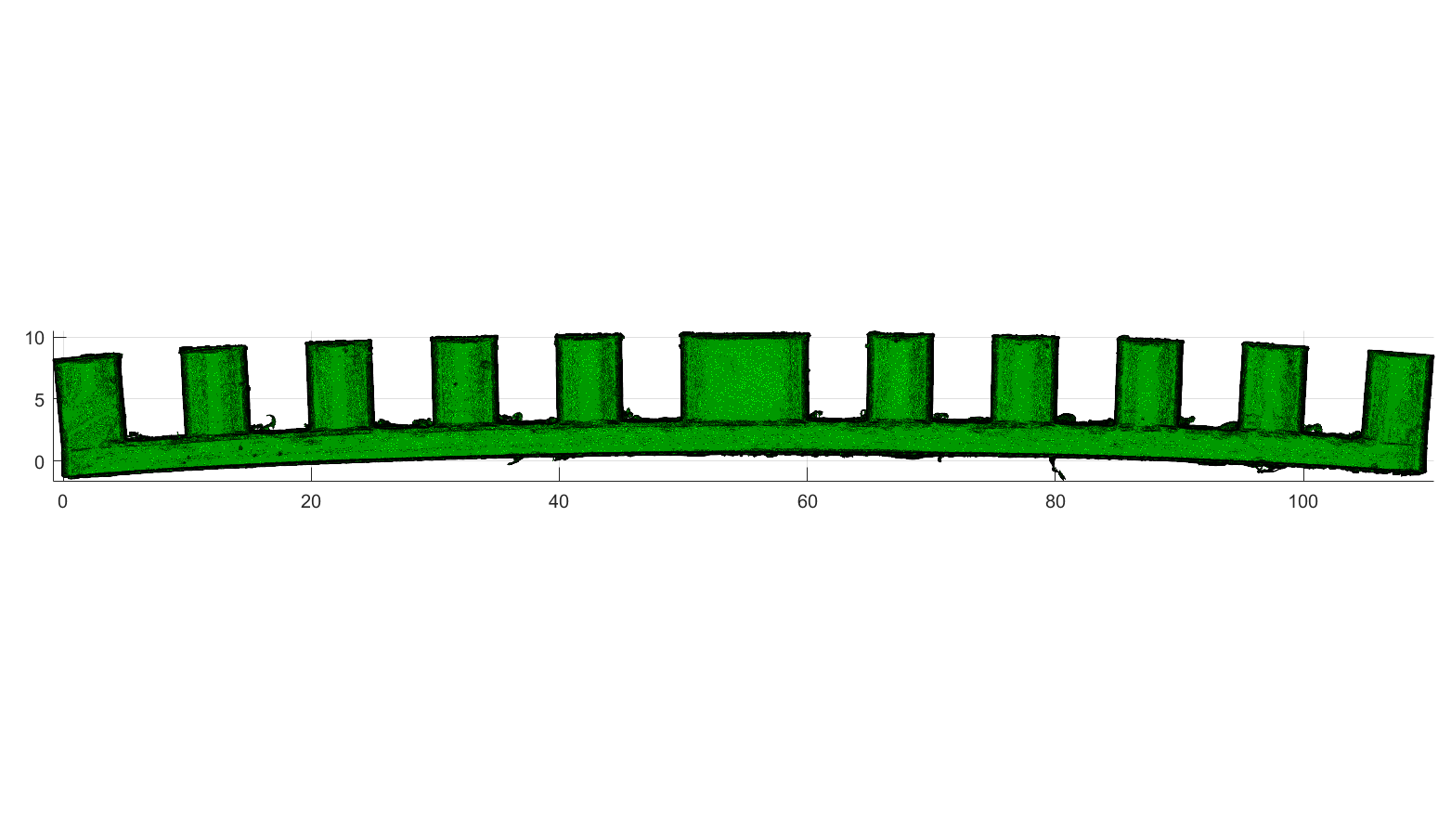}
    \caption{Front view.}
  \end{subfigure}
  \caption{Spatial mesh and spatial material nodes of scan data.}
  \label{fig:scan_data}
\end{figure}

A flowchart of the material distortion finding algorithm with data pipeline
is shown in Fig. \ref{FIG:Flowchart}.
The starting point of this algorithm is two STL files. One of these STL files is the design mesh illustrated
in Fig. \ref{fig:design_data}, and another one is the scan mesh shown in Fig. \ref{fig:scan_data}.


\begin{figure}[!tbp]
  \begin{subfigure}[b]{0.45\textwidth}
    \includegraphics[width=\textwidth]{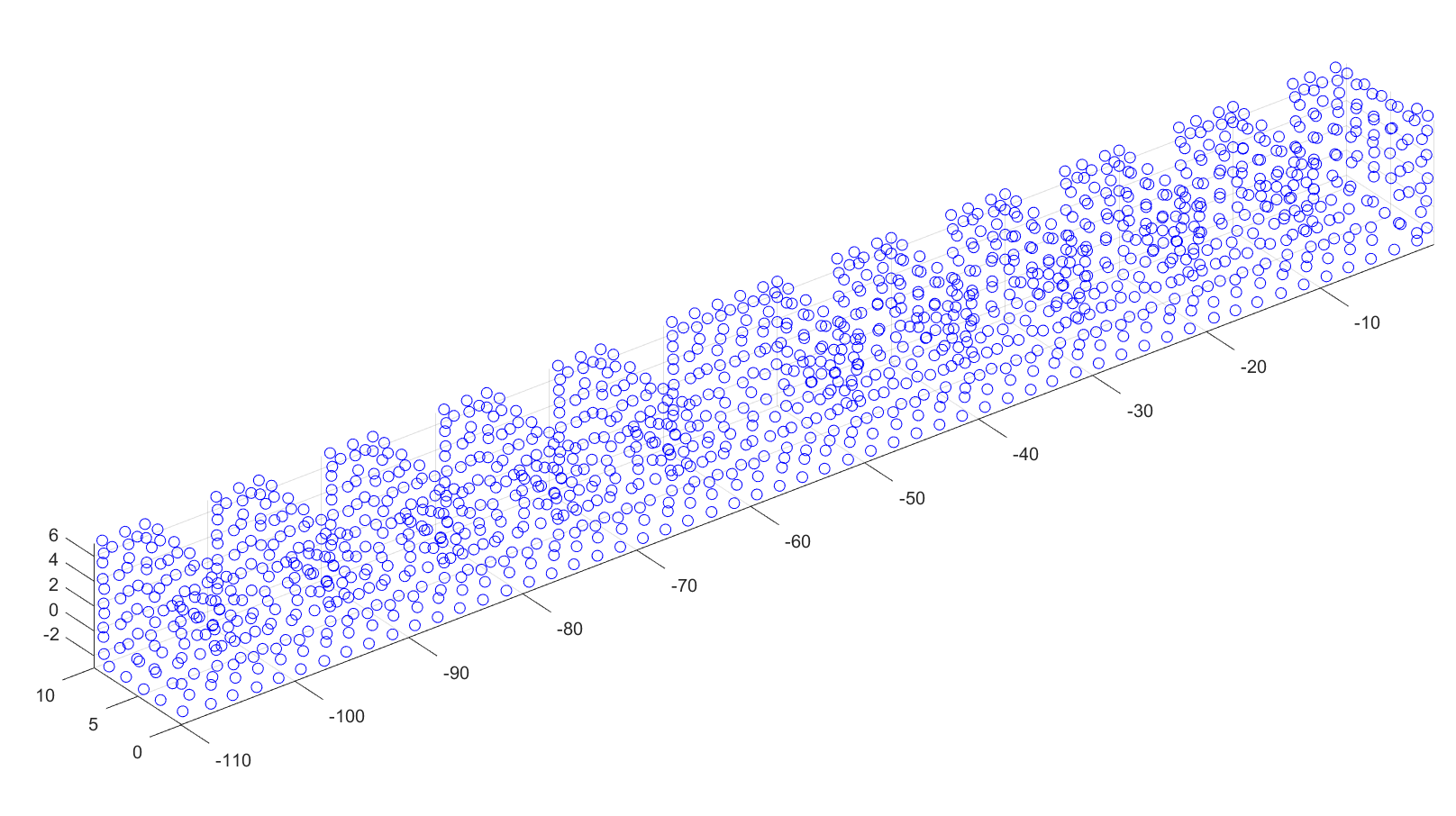}
    \caption{3D view.}
  \end{subfigure}
  \hfill
  \begin{subfigure}[b]{0.45\textwidth}
    \includegraphics[width=\textwidth]{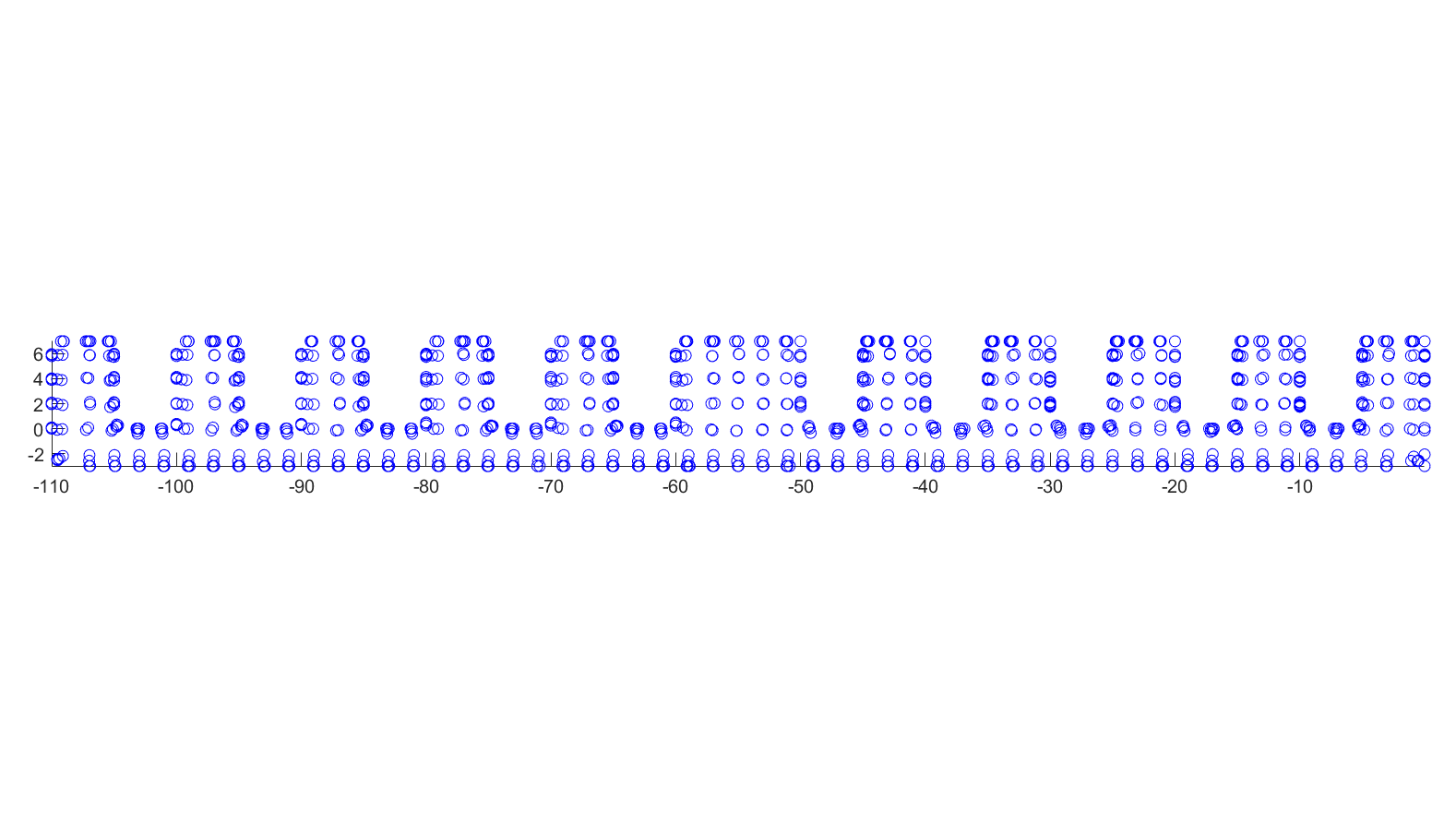}
    \caption{Front view.}
  \end{subfigure}
  \caption{point set of design data after downsampling.}
  \label{fig:design_data_down}
\end{figure}

\begin{figure}[!tbp]
  \begin{subfigure}[b]{0.45\textwidth}
    \includegraphics[width=\textwidth]{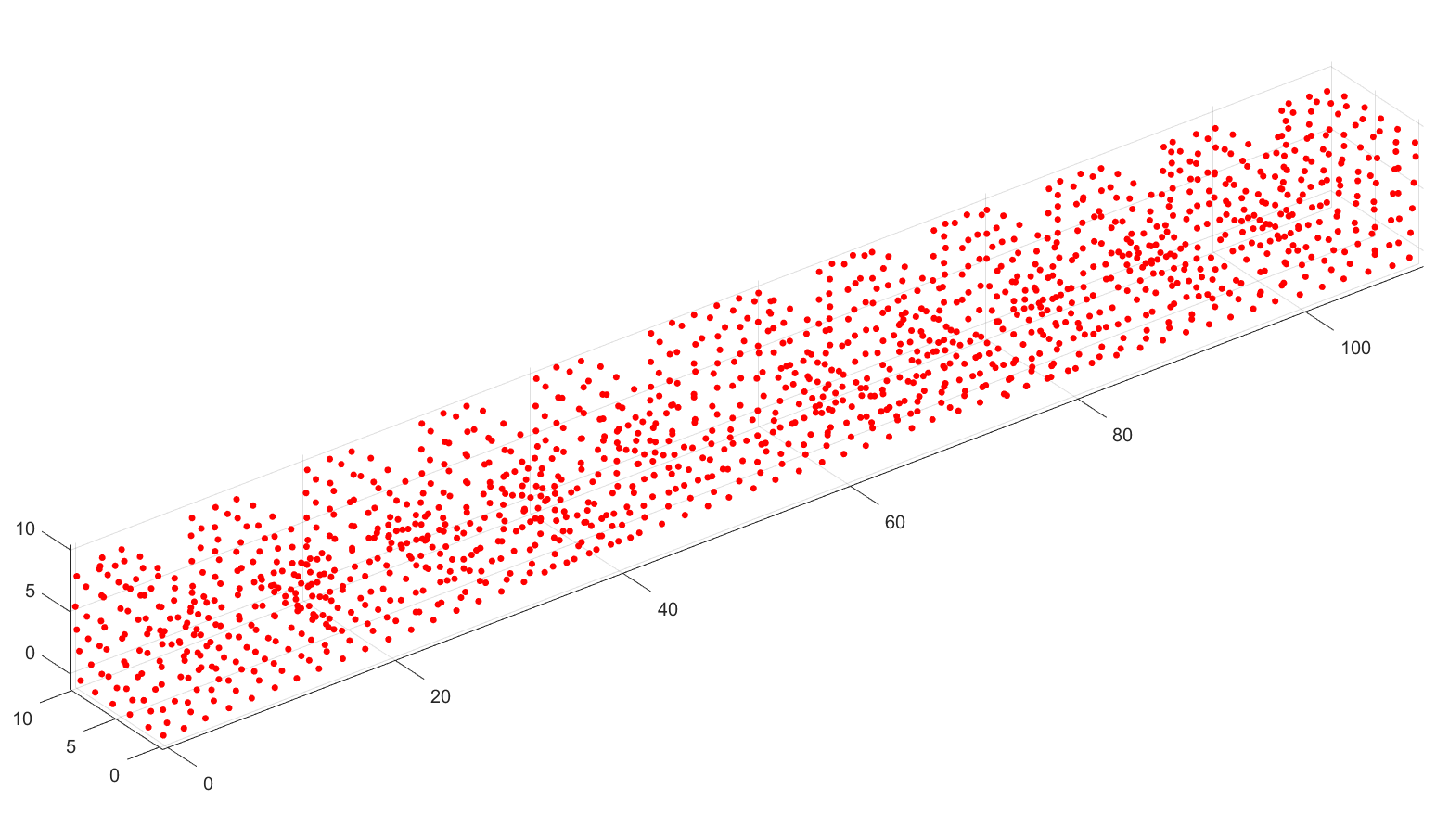}
    \caption{3D view.}
  \end{subfigure}
  \hfill
  \begin{subfigure}[b]{0.45\textwidth}
    \includegraphics[width=\textwidth]{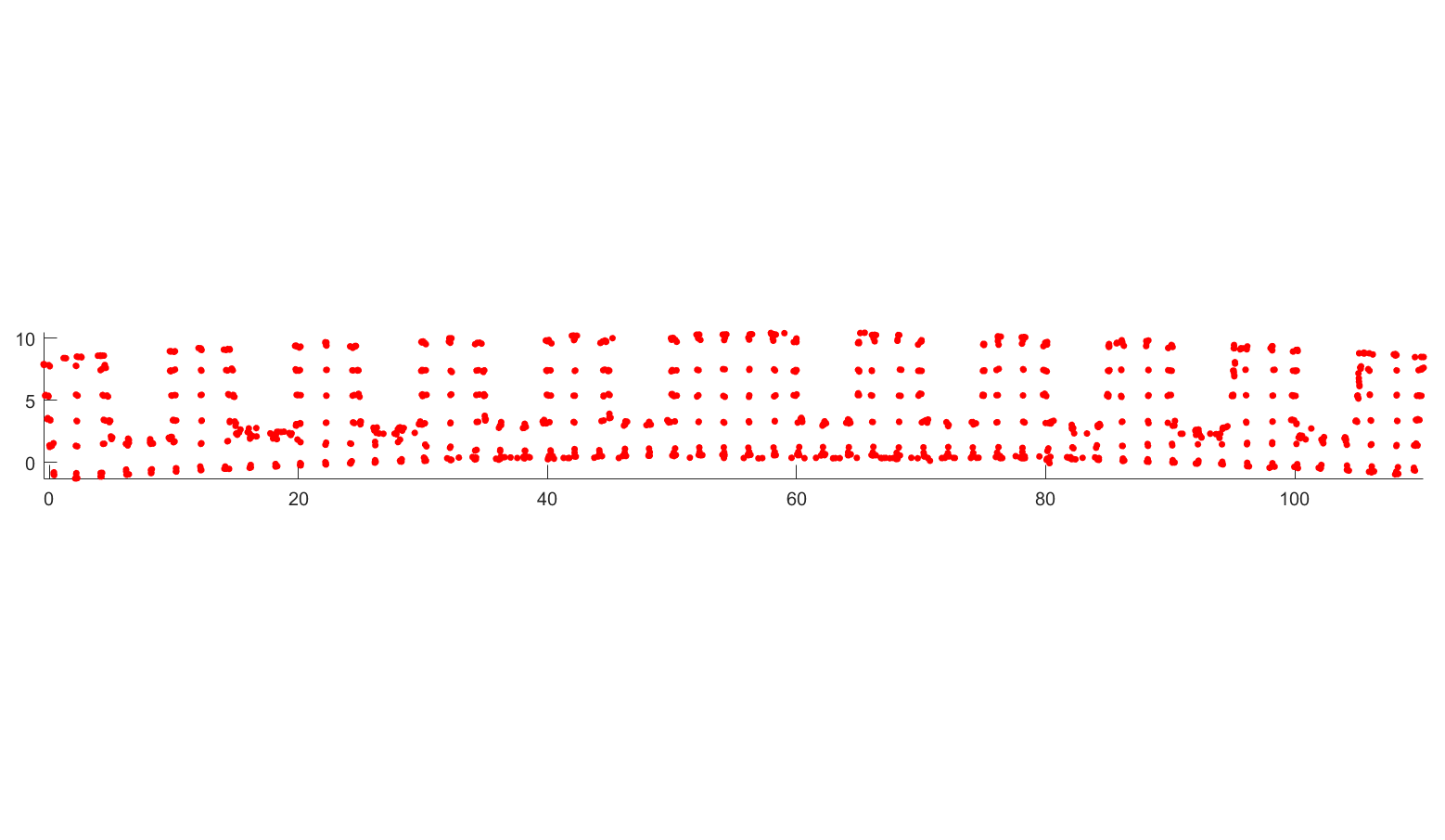}
    \caption{Front view.}
  \end{subfigure}
  \caption{point set of scan data after downsampling.}
  \label{fig:scan_data_down}
\end{figure}

Following the flowchart of the data pipeline,
the next step is to create two feature point sets for the material deformation finding algorithm.
Intuitively, we can choose all the nodes in two STL meshes as the feature points.
Nevertheless, as discussed in the previous subsection,
we only know spatial positions of material points
in the current configuration.
However, we do not know their corresponding material coordinates in the referential configuration.
In this particular case,
there are 1,990,267 nodes in scan mesh and 44,868 nodes in design mesh,
which means that it will be time-consuming to align all the original nodes directly.
Therefore, in order to mitigate the above problem, we downsample the nodes by using a box grid filter.
The size of the box grid filter is chosen as 2 mm in each direction
for both meshes, which means that we enforce the condition
that there is only one point in a size $2 mm^3$ cube.
If there are more than one nodes in a size $2 mm^3$ cube,
we will randomly choose one of the points in that cube.
The results of the downsampled data are shown in Fig. \ref{fig:design_data_down} and Fig. \ref{fig:scan_data_down}.


\begin{figure}[!tbp]
  \begin{subfigure}[b]{0.45\textwidth}
    \includegraphics[width=\textwidth]{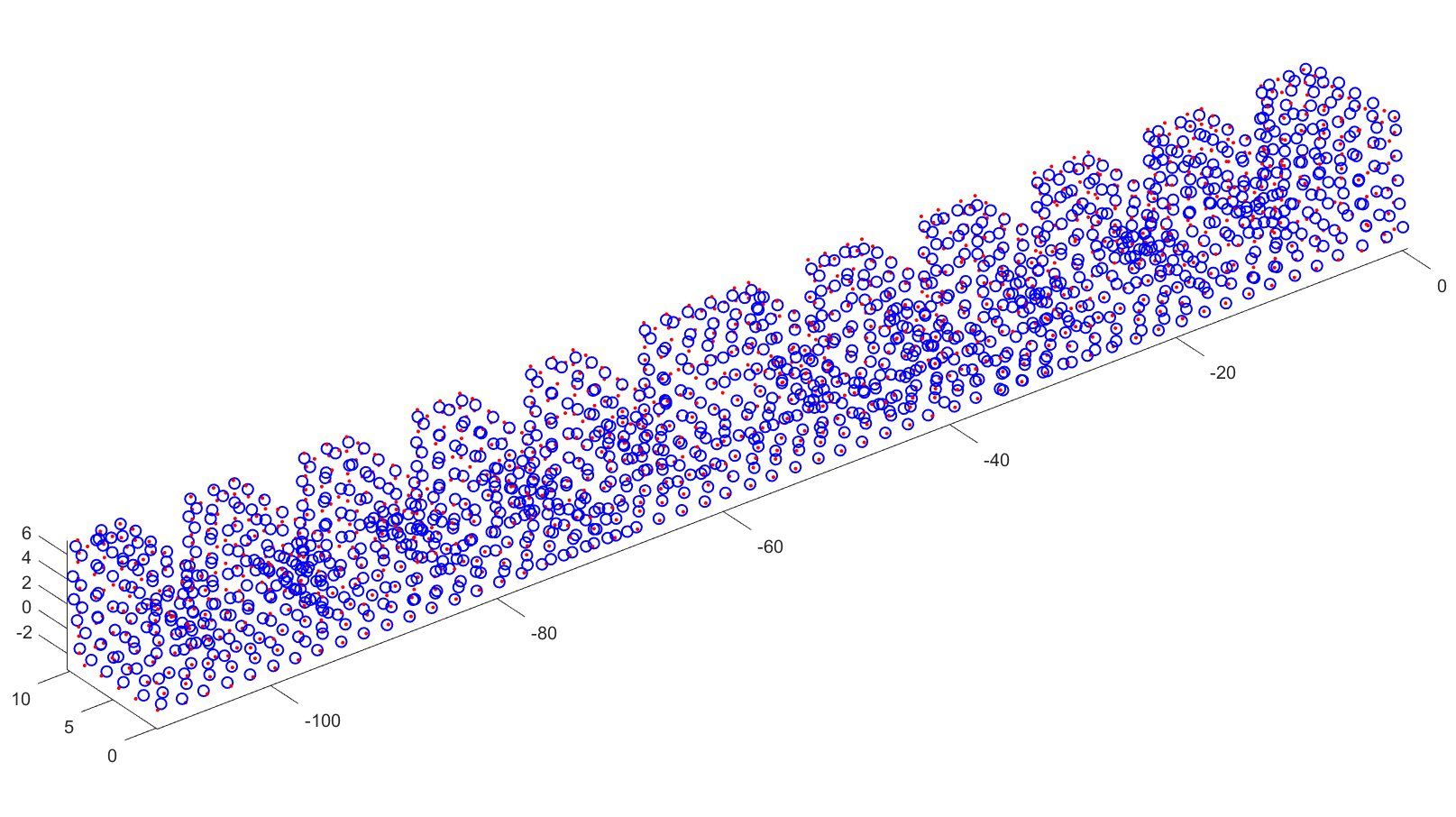}
    \caption{3D view.}
  \end{subfigure}
  \hfill
  \begin{subfigure}[b]{0.45\textwidth}
    \includegraphics[width=\textwidth]{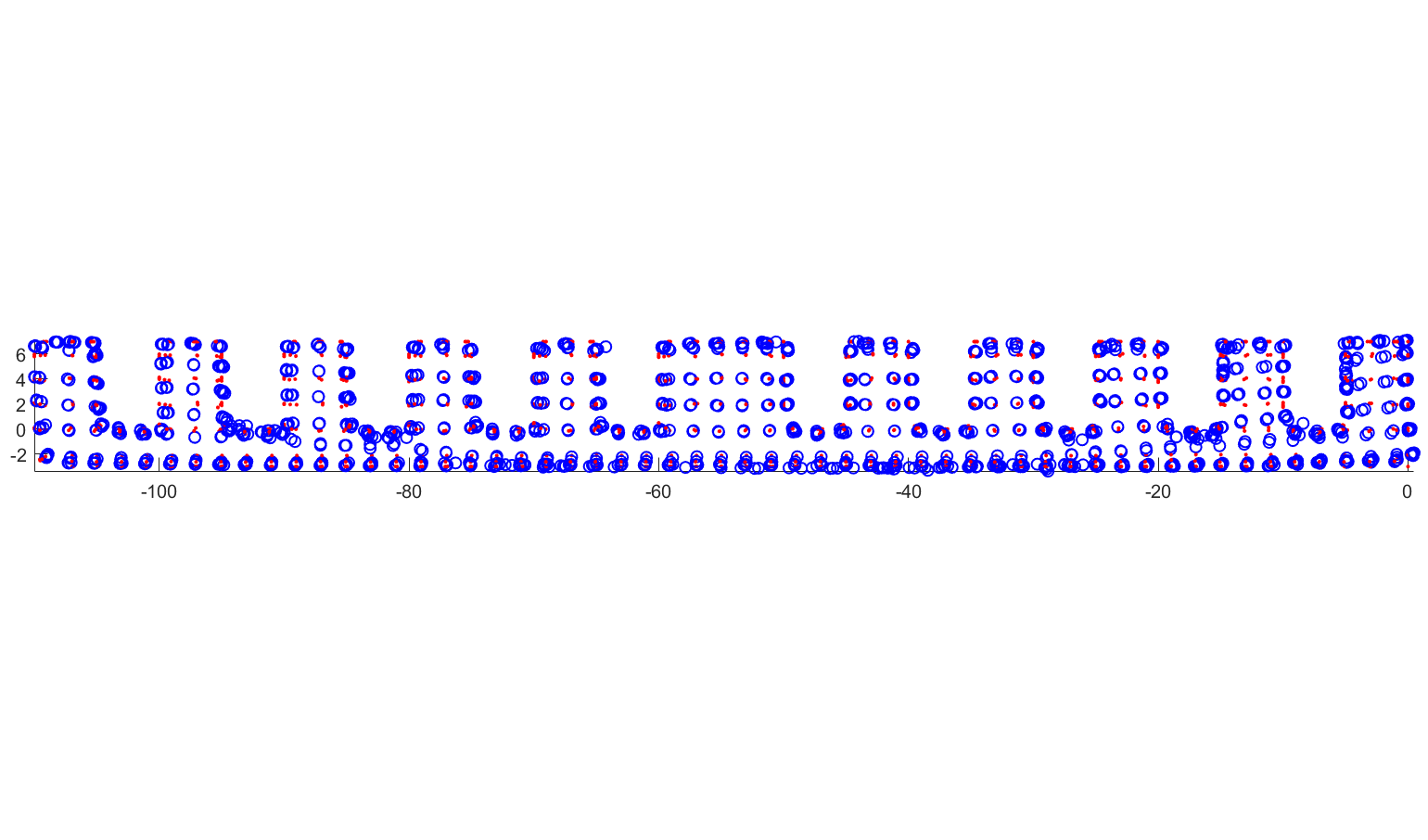}
    \caption{Front view.}
  \end{subfigure}
  \caption{Registration result of two point sets.}
  \label{fig:align_result}
\end{figure}

In the next step, we start to use the material deformation finding algorithm
to register two downsampled feature point sets with chosen parameters.
 In the simulation example reported in this paper,
 we choose the following values of the parameters:
 $ \beta=3, \alpha =2$, and $w =0.1$.
 In this example, the scan feature points are treated as GMM centroid,
 and design feature points are regard as the data points.
 Therefore, ideally, the scan feature points will move coherently to match
 the design feature points,
 and the displacement field can be interpolated by the displacements of the GMM centroid points.
The registration result is illustrated in Fig. \ref{fig:align_result}.
The result of this registration shows that most of the feature points move coherently
to the correct direction. However, there are still a few feature points
that are misled by the noise in the data and could not find the correct corresponding positions.

To resolve this issue, we need to clean the GMM centroid displacements
outlier before using them to generate the displacement field for the test specimen model.
An robust outlier detection algorithm
\cite{duncan2010universal} are used to get rid of outliers and smooth the displacement interpolation field.
The idea of this robust outlier detector is: the fluctuation of the of a given vector field is measured by a normalized residual fluctuation in its $k$ nearest neighbors \cite{westerweel2005universal}.
In this method, the normalized residual is defined in the following equation,
\begin{equation}
r_{0}^{*}=\frac{\left|\frac{U_{0}}{\operatorname{med}\left(d_{i}\right)
+\varepsilon_{a}}-\operatorname{med}\left(\frac{U_{i}}{d_{i}
+\varepsilon_{a}}\right)\right|}{\operatorname{med}\left|\frac{U_{i}}{d_{i}
+\varepsilon_{a}}-\operatorname{med}\left(\frac{U_{i}}{d_{i}+\varepsilon_{a}}\right)\right|+\varepsilon_{a}}
\label{eq:outlier_detector}
\end{equation}
where $U_{0}$ is the given vector value measured at one  data point;
$U_{i}$ is the given vector value of its neighbors;
$\varepsilon_{a}$ is the tolerance,  and $d_{i}$ is the distance from the data point to its $i$-th neighbor.
After calculating the normalized residual $r_{0}^{*}$,
it is compared with a threshold (default value is 2).
If the value of $r_{0}^{\prime}$
is greater than the threshold value, this data point is marked
as an outlier, and it should not be used in further calculation.

After detecting the outliers,
we can calculate the displacement field  $\mathbf{u}$ by cleaned GMM centroid displacements
\begin{equation}
    \mathbf{u} = \mathbf{Y}^{\prime} - \mathbf{y}
\end{equation}
where $\mathbf{y}$ is the spatial location of the scan data nodes,
$\mathbf{Y}^{\prime}$ is the location of the nodes of the scan data after material deformation finding,
which, as shown above, is a good approximation of ${\bf Y} \approx {\bf Y}^{\prime}$.
Then, we can use the displacement value at GMM centroid to interpolate the displacements
of other nodes. Both simple linear interpolation and $k$-nearest neighbors (KNN)
regression \cite{altman1992introduction} are used for the interpolation.
KNN regression is a non-parametric method,
which interpolates the displacement field in a new configuration
by its $k$ closest training data in the feature space.
Let $N_{k}(\mathbf{y})$ denote the $k$ values of $\mathbf{y}_{1}, \ldots, \mathbf{y}_{k}$
that are closest to a new location $\mathbf{y}$ where need to interpolate.
Then the estimator is just the mean over the corresponding displacement value $\mathbf{u}_i$ values,
\begin{equation}
\mathbf{u}(\mathbf{y})=\frac{1}{k} \sum_{i: \mathbf{y}_{i} \in N_{k}(\mathbf{y})} \mathbf{u}_{i}~.
\end{equation}
The motivation for using this estimator is that nearby observations is representative
for the samples around a particular
point $\mathbf{y}$ \cite{dudani1976distance}.
Although this is usually not true, it may be a good approximation,
if $\mathbf{u}$ is smooth and the neighborhood support is small.

After finding the thermal induced displacement field, many compensation principles can be used to redesign the model.
The simplest but still efficient way is to add an inverse displacement field to the original design mesh. The new added inverse displacement field will counteract the thermal induced distortion and finally achieve an neutral shape as prospect.

\begin{figure}[!tbp]
  \begin{subfigure}[b]{0.48\textwidth}
    \includegraphics[width=2.5in]{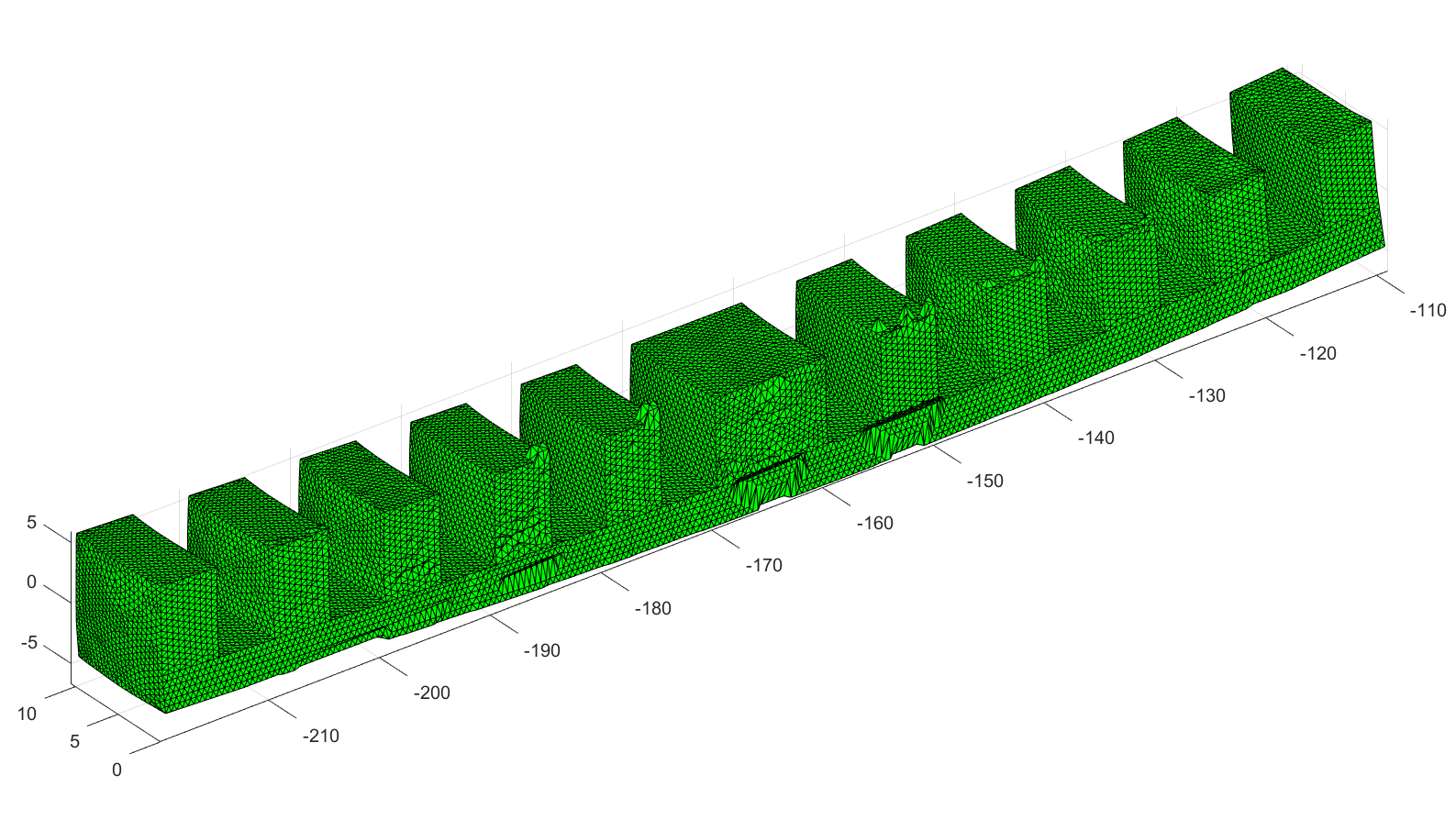}
    \caption{3D view.}
  \end{subfigure}
  \hfill
  \begin{subfigure}[b]{0.48\textwidth}
    \includegraphics[width=2.5in]{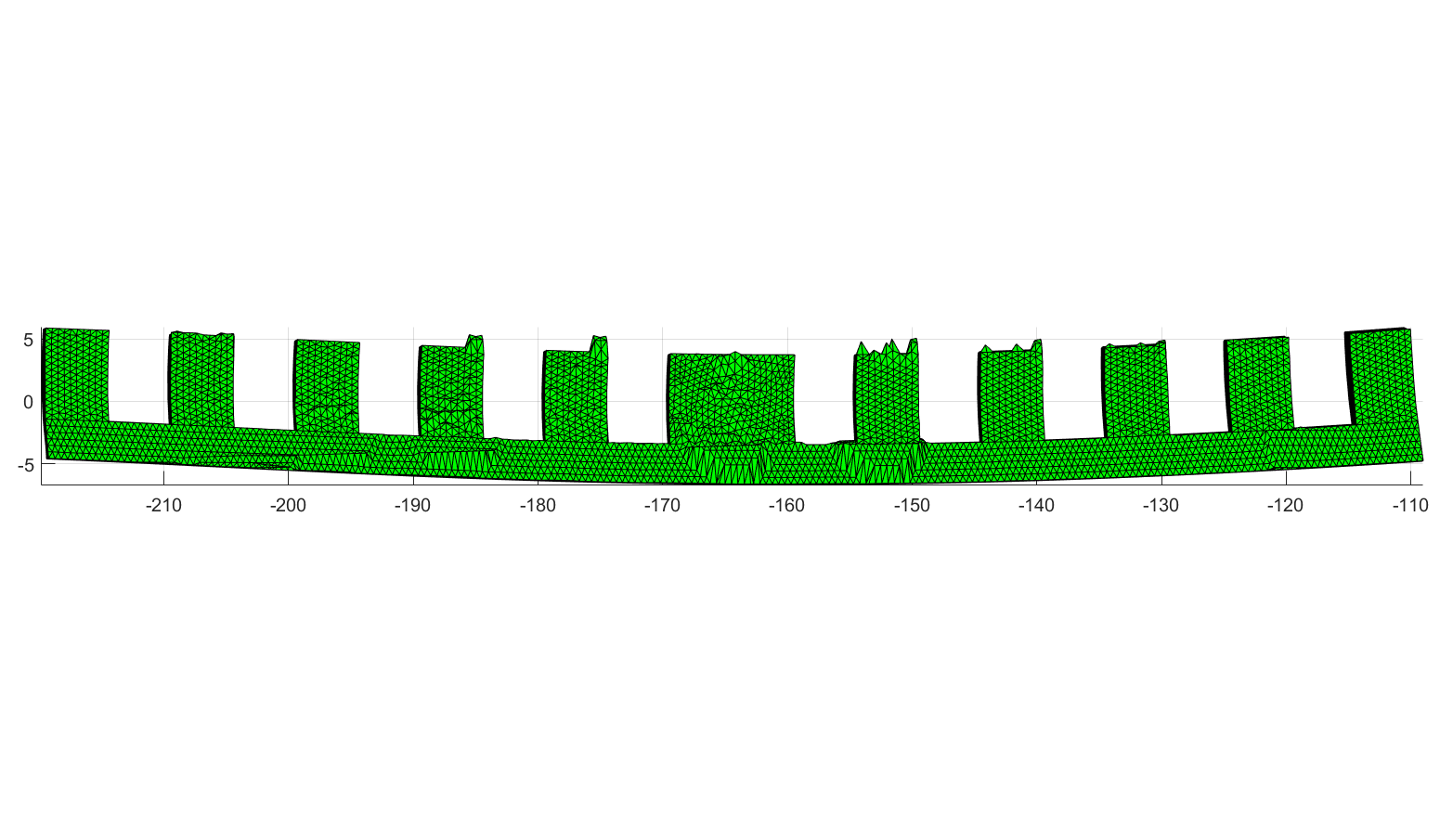}
    \caption{Front view.}
  \end{subfigure}
  \caption{Redesigned model with linear interpolation.}
  \label{fig:linear_interpolation}
\end{figure}

\begin{figure}[!tbp]
  \begin{subfigure}[b]{0.48\textwidth}
    \includegraphics[width=2.5in]{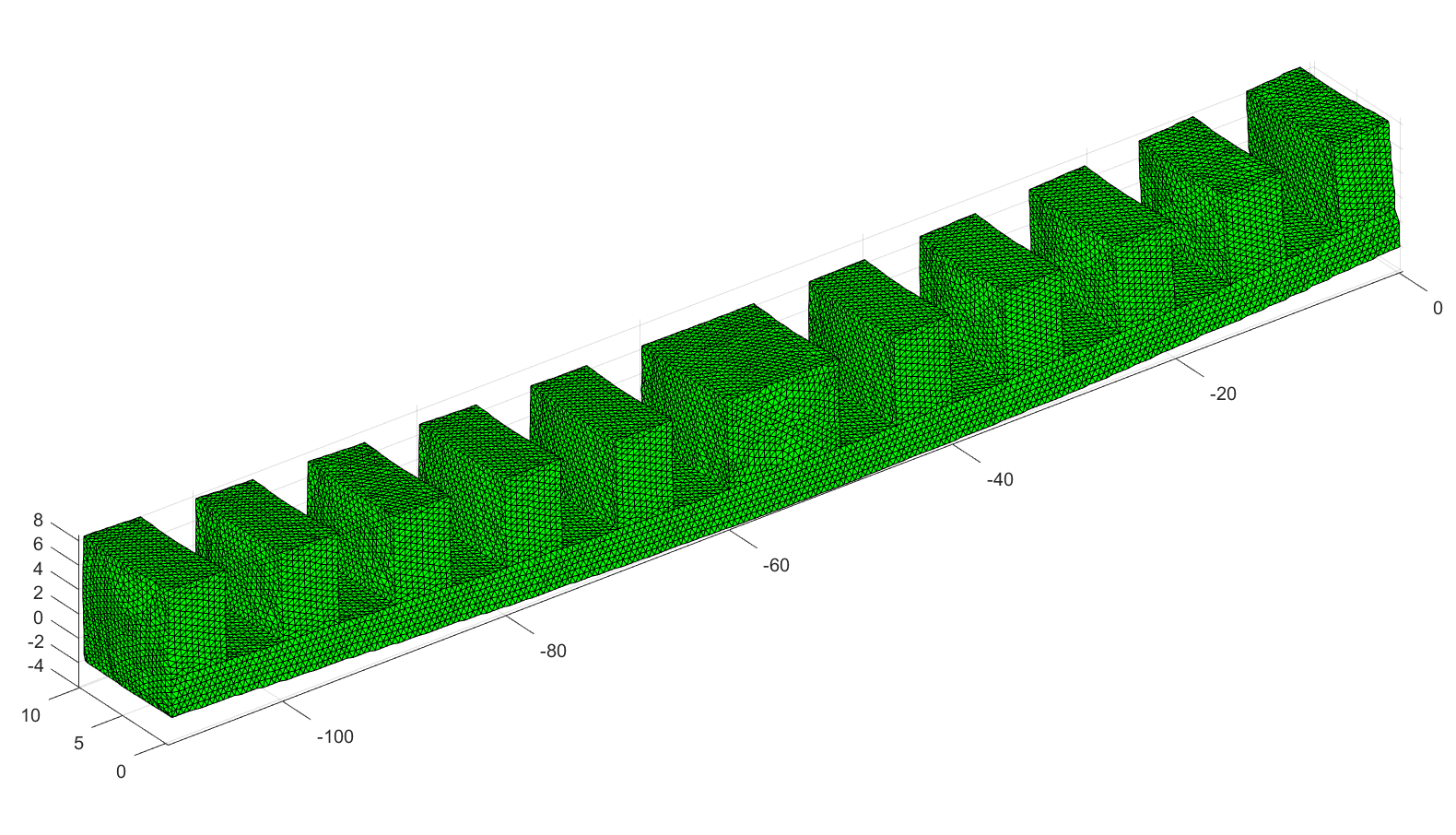}
    \caption{3D view.}
  \end{subfigure}
  \hfill
  \begin{subfigure}[b]{0.48\textwidth}
    \includegraphics[width=2.5in]{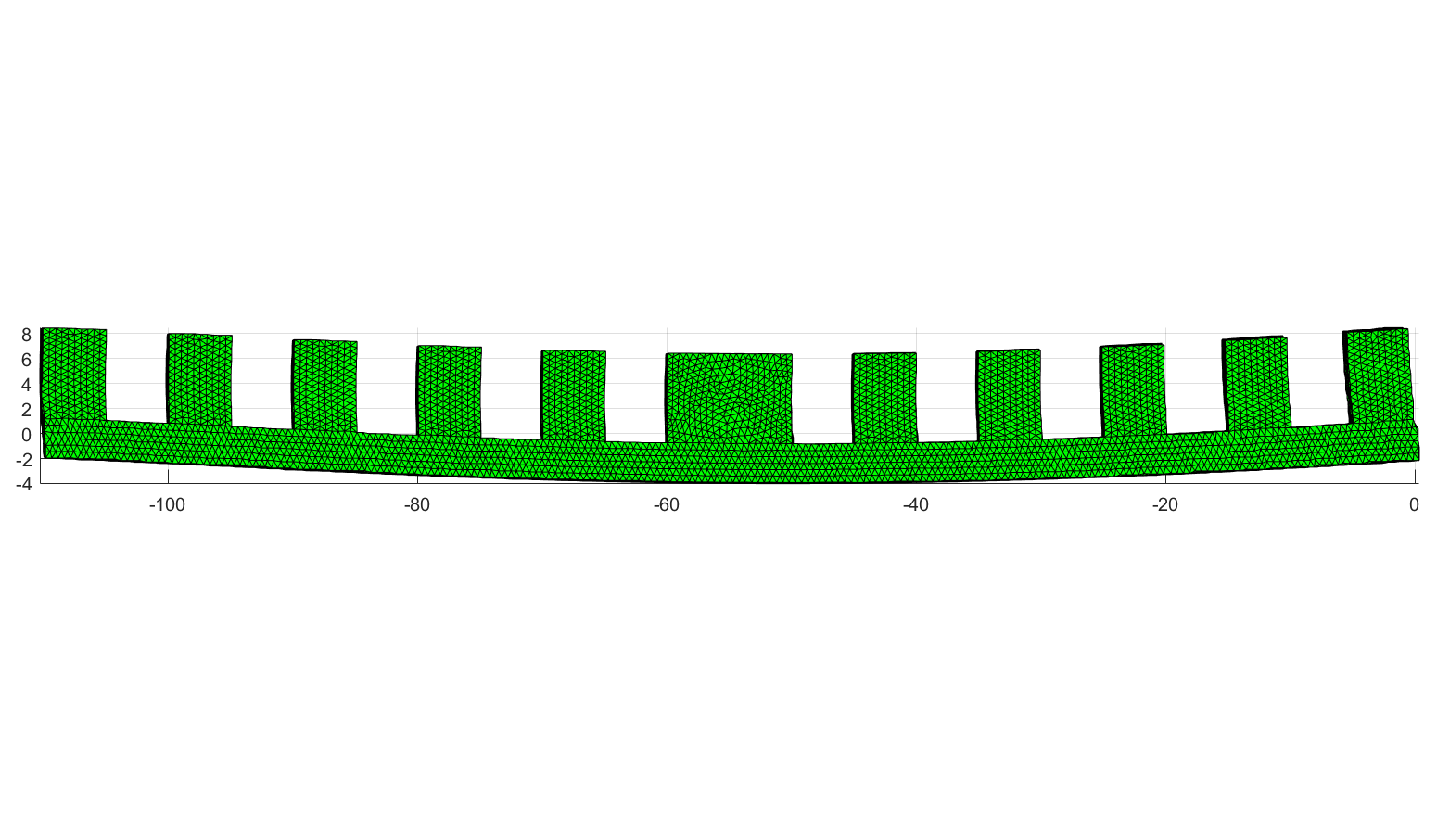}
    \caption{Front view.}
  \end{subfigure}
  \caption{Redesigned model with KNN Regression.}
  \label{fig:KNN_interpolation}
\end{figure}

Both redesigned results of linear interpolation and KNN regression are illustrate in Fig. \ref{fig:linear_interpolation} and Fig. \ref{fig:KNN_interpolation}.
The results of both redesigned mesh indicate that KNN regression is significantly better than the simply linear regression.
These differences can be explained in part by the robustness of KNN regression,
which does not require the regression function passing through all the data points.

In summary,
the new thermal deformation compensation method uses material deformation finding algorithm to align feature points which are nodes in triangular meshes. Then after removing the outlier in displacement field, KNN regression is used to interpolate the displacement field at other points.
Finally, redesign the model to compensate the distortion.
This new compensation method starts with scanned data and design data files and ends up with a new redesigned STL file which could be use to print.
It defines a new design-cycle to improve the quality of final produced 3D printing models.
Future improvements may include design
a more intelligent and maybe iterative way to use the thermal induced deformation field to redesign the model.

\section{Conclusions}
In this paper, we introduce a material distortion finding
algorithm for predicting thermal compensation allowance
in additive manufacturing.

In this data-driven approach,
we only need to know the spatial positions of material points, i.e. the scan position
in the current configuration,
but not the corresponding material coordinates in the referential configuration or the designed configuration.
By fully using the information contained in STL files,
a unique weight associated with each feature point are calculated based on the area of triangle elements in STL files,
which will lead to the spatial material points converge back to its original or material coordinates based
on the maximum likelihood optimization.
Finally, one can find the displacements as well as deformation of all the material points contained in the scan data.
Different numerical tests have been carried out,
and comparison study of results are also conducted.
The material distortion finding algorithm developed here is illustrated to offer robustness to a greater degree with accuracy through quantitative evaluations,
and it paved the way for registration to continuum manifold for significant applications
in 3D printing technology.

\section*{Acknowledgements}
S. L. would like to thank Ford Motor Company for awarding a Ford-University grant to support this research,
and C. W. would like to thank a graduate fellowship from Harbin Engineering University supporting
his Ph.D. study.



\bigskip
\bigskip
\section*{References}
\bibliographystyle{cas-model2-names}

\bibliography{cas-refs}

\end{document}

%% file: macros.tex
\newcommand{\real}{\hbox{\rm I\kern-0.2emR}}

\newcommand{\integer}{\hbox{\rm I\kern-0.2emN}}

\newcommand{\Hhset}{\hbox{\rm I\kern-0.2emH}}
\newcommand{\Thset}{\hbox{\rm T\kern-0.3emI}}

\def\boxit#1{\vbox{\hrule\hbox{\vrule\kern7pt\vbox{\kern7pt
\hbox{$\displaystyle#1$}\kern7pt}\kern7pt\vrule}\hrule\kern-8.5pt}}
\newtheorem{remark}{Remark}[section]

\usepackage{float}